\documentclass[11pt]{article}

\usepackage{graphicx}
\usepackage{amssymb}
\usepackage{amsmath}
\usepackage{graphicx}
\usepackage{amstext}
\usepackage{esint}

\newcommand{\Omo}{\Omega_m^0}

\newcommand{\ORo}{\Omega_{R}^0}
\newcommand{\OL}{\Omega_{\Lambda}}

\newcommand{\OLo}{\Omega_{\Lambda}^0}

\newcommand{\rmo}{\rho_{m}^0}

\newcommand{\rmr}{\rho_m}

\newcommand{\rL}{\rho_{\CC}}

\newcommand{\rLo}{\rho_{\CC}^0}

\newcommand{\CC}{\Lambda}

\newcommand{\nueff}{\nu_{\rm eff}}
\newcommand{\nueffp}{\nu_{\rm eff}^{\prime}}

\newcommand{\xiR}{\xi'}
\newcommand{\rRo}{\rho_r^0}
\newcommand{\dH}{\dot{H}}
\newcommand{\alem}{\alpha_{\rm em}}

\newcommand{\be}{\begin{equation}}
\newcommand{\ee}{\end{equation}}

\newcommand{\LQCD}{\Lambda_{\rm QCD}}

\newcommand{\Oro}{\Omega_{r}^0}

\begin{document}

\hyphenation{cos-mo-lo-gists un-na-tu-ral-ly in-te-gra-ting
ne-gli-gi-ble e-xis-ten-ce con-vin-cing des-crip-tion ma-xi-mum}

\begin{center}

{\large \text{\bf COSMOLOGICAL CONSTANT VIS-\`A-VIS DYNAMICAL VACUUM:}} {\large \text{\bf BOLD CHALLENGING THE $\CC$CDM}\footnote{Invited review paper based on the invited plenary talk at the International Conference on New Physics at the Large Hadron Collider, NTU, Singapore, 2016. Some of the results and discussions  presented here have been further updated and expanded with respect to the published version.}} \vskip 2mm

 \vskip 10mm

\textbf{\large Joan Sol\`a}

\vskip 0.5cm

Departament de F\'\i sica Qu\`antica i Astrof\'\i sica ,\\
Av. Diagonal 647, E-08028 Barcelona, Catalonia, Spain

\vskip0.15cm
and
\vskip0.15cm

Institute of Cosmos Sciences (ICCUB)\\
Univ. de Barcelona, Av. Diagonal 647, E-08028 Barcelona

\vskip0.4cm

E-mail:    sola@fqa.ub.edu
 \vskip2mm

\end{center}
\vskip 15mm

\begin{quotation}
\noindent {\large\it \underline{Abstract}}.
Next year we will celebrate 100 years of the cosmological term, $\CC$, in Einstein's gravitational field equations, also 50 years since the cosmological constant problem was first formulated by Zeldovich, and almost about two decades of the observational evidence that a non-vanishing, positive, $\CC$-term could be the simplest phenomenological explanation for the observed acceleration of the Universe. This mixed state of affairs already shows that we do no currently understand the theoretical nature of $\CC$. In particular, we are still facing the crucial question whether $\CC$ is truly a fundamental constant or a mildly evolving dynamical variable. At this point the matter should be settled once more empirically and, amazingly enough, the wealth of observational data at our disposal can presently shed true light on it. In this short review I summarize the situation of some of these studies. It turns out that the $\CC=$const. hypothesis, despite being the simplest, may well not be the most favored one when we put it in hard-fought competition with specific dynamical models of the vacuum energy. Recently it has been shown that the overall fit to the cosmological observables SNIa+BAO+$H(z)$+LSS+BBN+CMB do favor the class of ``running'' vacuum models (RVM's) -- in which $\CC=\CC(H)$ is a function of the Hubble rate -- against the ``concordance'' $\CC$CDM  model. The support is at an unprecedented level of $\gtrsim4\sigma$ and is backed up with Akaike and Bayesian  criteria leading to compelling evidence in favor of the RVM option and other related dynamical vacuum models. I also address the implications of this framework on the possible time evolution of the fundamental constants of Nature.

\vspace{0.5cm}


\end{quotation}

\newpage

\tableofcontents

\newpage

\vskip 6mm

\section{Introduction}\label{sect:introduction}

On February 15th 1917 the famous seminal paper where Einstein introduced the cosmological term $\CC$ (actually denoted ``$\lambda$'' in it)  along with the fundamental observation that such term does not destroy the general covariance of his original (1915) field equations, was issued \footnote{``Wir k\"onnen n\"amlich auf der linken Seite der Feldgleichungen [(13)] den mit einer vorl\"aufig unbekannten universellen Konstante -$\lambda$ multiplizierten Fundamentaltensor $g_{\mu\nu}$ hinzuf\"ugen, ohne da{\ss} dadurch die allgemeine Kovarianz zerst\"ort wird...'' (A. Einstein\,\cite{Einstein1917})}. Since then, and despite its lengthy and tortuous history, $\CC$ has traditionally been associated to the concept of vacuum energy density: $\rL = \CC/(8\pi G)$, where $G$ is Newton's constant. This association is somehow natural if we take into account that the vacuum energy is conceived as being uniformly distributed in all corners of space. In other words, it perfectly preserves the Cosmological Principle. Thus it cannot be related to any form of matter content of the Universe, which always tends to cluster through gravitational collapse. However, it is difficult to understand the origin of the vacuum energy (what is it after all?) unless one makes a connection with quantum physics, where the vacuum fluctuations of the fields are part of everyday's life. Such connection was first pointed out by Zeldovich in 1967 \,\cite{Zeldovich1967}.

Unfortunately, while the vacuum fluctuations in the presence of real particles (external lines of Feynman diagrams) are well understood in quantum field theory (QFT) -- they are called ``radiative corrections'' to the classical processes ---, the pure vacuum-to-vacuum diagrams, i.e. closed diagrams without external legs, are not so well understood and find themselves in a peculiar situation. They cannot be considered radiative ``corrections''  because they are not related to any classical or tree-level amplitude to be ``corrected''! In other words, they are pure quantum effects with no classical counterpart. In both cases we are dealing with quantum effects, and as such they are usually UV-divergent and require regularization and renormalization procedures. In the case of the radiative corrections, the final result is always a small finite contribution to the classical quantity with which we started the zeroth order computation  (e.g. a cross-section, decay rate, energy level etc). In contrast, if we attempt to renormalize a zero-point energy (ZPE) diagram with no external legs, i.e. a vacuum-to-vacuum diagram, we find that the renormalization procedure despite being perfectly possible as in the radiative calculation case,  the resulting finite contribution is always much larger (by many orders of magnitude) to the measured value of $\rL$\,\cite{SNold,SNmod,WMAP9-2013,PLANCK2013,PLANCK2015}. And this is so no matter what is the sort of (boson or fermion) field we are considering.  For example, take the standard model (SM) of strong and electroweak interactions. Let
$M_H\simeq 125$ CeV be the (measured) Higgs mass and $M_F\equiv
G_F^{-1/2}\simeq 293$ GeV the Fermi scale. The contribution to the
ZPE from the Higgs field is of order $M_H^4\sim 10^8$ GeV$^4$, and
the ground state value of the (classical)  Higgs potential reads
$\langle V\rangle=-(1/8\sqrt2) M_H^2\,M_F^2\sim -10^9$ GeV$^4$. In
magnitude this is of order $\sim v^4$, where $v\simeq 246$ GeV is the
vacuum expectation value of the scalar field. Equally significant is
the ZPE from the top quark (with mass $m_t\simeq 174$ GeV), which is
of order  $m_t^4\sim 10^9$ GeV$^4$ and negative (because it is a
fermion). After adding up all these effects a result of the same
order ensues which is $\left(10^9/10^{-47}\right)\sim 10^{56}$ times
bigger than what is needed. It is senseless to expect that these contributions may cancel, all the more so bearing in mind that they must be retuned order by order in perturbation theory!

The above situation is of course (part of) the so-called ``old cosmological constant problem''\,\cite{CCproblem1,CCproblem2} -- see also \cite{JSPReview2013,SolaGomez2015}.
As mentioned, this problem was first formulated by Zeldovich about half a century ago\,\cite{Zeldovich1967} and is the main source of headache for every theoretical cosmologist confronting his/her theories with the measured value of $\rL$. Furthermore, the purported discovery of the Higgs boson at the LHC has accentuated the CC problem greatly, certainly much more than is usually recognized.
Owing to the necessary spontaneous symmetry breaking (SSB) of the electroweak (EW) theory, an induced contribution  to $\rL$ is generated which is appallingly much larger (viz. $\sim 10^{56}$) than the tiny value $\rL\sim 10^{-47}$ GeV$^4$ (``tiny'' only within the particle physics standards, of course) extracted from observations. So the world-wide celebrated  ``success'' of the Higgs finding  in particle physics actually became a cosmological fiasco, since it instantly detonated the ``modern CC problem'', i.e. the confirmed size of the EW vacuum, which should be deemed  as literally ``real'' (in contrast to other alleged -- ultralarge -- contributions from QFT)  or ``unreal'' as the Higgs boson itself! One cannot exist without the other.

In this work we will not further address the formal QFT issues mentioned above, as they go beyond the scope of the kind of presentation adopted here, which is essentially phenomenological. I refer the reader to review works, e.g. \,\cite{CCproblem1,CCproblem2,JSPReview2013,SolaGomez2015}, for a more detailed exposition of the CC problem and the general \textit{dark energy} (DE) problem\,\cite{DEBook}. 
Setting aside the ``impossible'' task of predicting the $\CC$ value itself -- unless it is understood as a ``primordial renormalization''\,\cite{GRF2015} -- I will focus here on a more phenomenological presentation, which may help to decide if the $\CC$-term is a rigid cosmological constant or a mildly evolving dynamical variable. In the latter case there would be more room for considering $\CC$ as a  more sophisticated QFT object within the kind of scenarios that I will describe here. Having this purpose in mind I will review several types of dynamical vacuum models (DVM's) and shall put special emphasis on a subclass of them in which $\CC$ appears neither as a rigid constant nor as a scalar field (quintessence and the like)\,\cite{CCproblem2}, but as a ``running'' quantity in QFT in curved spacetime. This is a natural option for an expanding Universe. As we will show, such kind of dynamical vacuum models are phenomenologically quite successful; in fact so successful that they are currently challenging the $\CC$CDM\,\cite{ApJ2015,JCAP2015b,FirstEvidence2016, Comparison2016,PRDcompanion,MNRAS2015,JCAP2015a,BPS2009,Grande2011}.

The layout of the paper is as follows. In Sect. 2 we introduce the general dynamical vacuum models (DVM's), which in some cases implies the dynamical evolution of the gravitational coupling and/or the anomalous conservation of matter. In Sect. 3 we present the running vacuum models (RVM's) and other related models. These are more specific DVM's and some of them have recently been shown to fir the overall cosmological data substantially better than the $\CC$CDM. The results of the fitting procedure are presented and discussed in Sect. 4. The possible impact of these successful dynamical vacuum models for a possible explanation of the the cosmic time evolution of masses and couplings (in particular the fine structure constant) is addressed in Sect. 5. Finally, in Sect. 6 we deliver our conclusions.

\section{Dynamical vacuum models (DVM's)}

The traditional hypothesis  $\CC=$const. for the $\CC$-term in Einstein's field equations
\begin{equation}
G_{\mu\nu}-\CC\,g_{\mu\nu}\equiv R_{\mu \nu }-\frac{1}{2}g_{\mu \nu }R-\CC\,g_{\mu\nu}=8\pi G\ {T}_{\mu\nu}\,,
\label{EEsplit}
\end{equation}
is the simplest one compatible with the Bianchi identity
$\nabla^{\mu}G_{\mu\nu}=0$ satisfied by the Einstein tensor on the \textit{l.h.s.} of Eq.\,(\ref{EEsplit}), provided one assumes local covariant matter conservation, $\nabla^{\mu}\,T_{\mu\nu}=0$, \emph{and} $G=$const. (i.e. rigid gravitational coupling). However, it is obvious that this scenario is just one particular set of assumptions among many other possibilities. If we define the vacuum energy density in the usual way, $\rL = \CC/(8\pi G)$, and move it to the \textit{r.h.s.} of Einstein's equations, then the Bianchi identity $\nabla^{\mu}G_{\mu\nu}=0$
leads in general to the following generalized conservation law for
the product of $G$ times the full energy-momentum tensor $\tilde{T}_{\mu\nu}$ on its \textit{r.h.s.} (i.e. the energy-momentum tensor after we include in it the vacuum energy density):
\begin{equation}\label{GBI}
\nabla^{\mu}\left(G\,\tilde{T}_{\mu\nu}\right)=\nabla^{\mu}\,\left[G\,(T_{\mu\nu}+g_{\mu\nu}\,\rL)\right]=0\,.
\end{equation}
Writing it out in the FLRW metric, we find:
\begin{equation}\label{BianchiGeneral1}
\sum_N\left[\frac{d}{dt}\,\left(G\,\rho_N\right)+3\,G\,H\,(\rho_N+p_N)\right]=0\,,
\end{equation}
where we have assumed the usual perfect fluid form for the matter energy-momentum tensor for all the fluid components $N$, characterized in each case by the density $\rho_N$ and pressure $p_N$ in the proper frame. These components receive contributions from matter (both relativistic and non-relativistic) as well as from the vacuum energy density, whose pressure satisfies $p_{\Lambda}=-\rho_{\Lambda}$. Therefore, we can be more explicit and write the previous equation as follows:
\begin{equation}\label{BianchiGeneral}
\frac{d}{dt}\,\left[G(\rho_m+\rho_r+\rL)\right]+3\,G\,H\,\sum_{i=m,r}(\rho_i+p_i)=0\,.
\end{equation}
Here $H=\dot{a}/a$ is the Hubble rate, $\rho_r$ denotes the relativistic (or radiation) component ($p_r=\rho_r/3$) and $\rmr$ the non-relativistic matter component (hence with $p_m=0$).
Equation (\ref{BianchiGeneral}) is actually a first integral of the basic system of equations emerging from the explicit form of Einstein's equations (\ref{EEsplit}) in the Friedmann-Lema\^\i tre-Robertson-Walker (FLRW) metric.  We will hereafter assume that the FLRW metric is characterized by spatially flat sections. This does not affect the structure of the conservation laws but can affect other equations, specially those related with the Hubble function and its time derivatives. In the flat FLRW case, Friedmann's equation together with the accelerating equation read
\begin{eqnarray}
&&3H^2=8\pi\,G\,(\rho_m+\rho_r+\rho_\Lambda)\label{eq:FriedmannEq}\\
&&3H^2+2\dot{H}=-8\pi\,G\,(p_r-\rho_\Lambda)\label{eq:PressureEq}\,.
\end{eqnarray}
They are equally valid if $G$ and/or $\rL$ are homogeneous and isotropic functions of the cosmic time or the scale factor, thereby evolving with the expansion. One can easily retrieve equation (\ref{BianchiGeneral}) from these two equations.

Remarkably, the local conservation law (\ref{BianchiGeneral}) mixes
the matter-radiation energy density with the vacuum energy $\rL$. In other words, the vacuum can decay into matter or the matter can decay into vacuum energy. This framework is perfectly compatible with the Cosmological Principle since the generalized conservation law was derived from the FLRW metric and hence respecting the postulate of homogeneity and isotropy.

Let us mention the following possibilities or cosmological vacuum types:
\begin{itemize}

\item Standard type or \textbf{$\CC$CDM}:  $G=$const. {and} $\rL=$const.:

If there are no other components in the cosmic fluid apart from relativistic and nonrelativistic matter, this is the
standard case or ``concordance'' $\CC$CDM cosmological model, implying the local covariant
conservation law of matter-radiation:

\begin{equation}\label{standardgeneralconserv}
 \dot{\rho}_m + 3H\rho_m+\dot{\rho}_r + 4H\rho_r =0.
\end{equation}
Except for the epoch near the equality of matter and radiation, these components are usually assumed to be conserved separately, hence
\begin{equation}\label{standardconserv}
\dot{\rho}_m+3\,H\,\rmr=0\,,
\end{equation}
and
\begin{equation}\label{standardconservRadiation}
\dot{\rho}_r+4\,H\,\rmr=0.
\end{equation}
Integration of these equations with respect to the scale factor variable $a(t)$ immediately leads to
\begin{eqnarray}
\rho_m &=& \rho_m^0 ~a^{-3} \label{standardlawmatter} \\
\rho_r &=& \rho_r^0 ~a^{-4}\,,\label{standardlawradiation}
\end{eqnarray}
where $\rho_m^0$ and $\rho_r^0$ are the respective current densities, i.e. the corresponding values at $a=1$.

\item  \textbf{Type-A model}: $G=$const {and} $\dot{\rho}_{\CC}\neq 0$:

Here Eq.(\ref{BianchiGeneral}) leads to an ``anomalous'' type of conservation law (this is what {\bf A} stands for):
\begin{equation}\label{mixed conslaw}
\sum_{i=m,r}\left[\dot{\rho}_i+3\,H\,(\rho_i+p_i)\right]=Q\,,
\end{equation}
in which there is a nonvanishing source  $Q=-\dot{\rho}_{\CC}$ causing the interaction of matter with the vacuum and therefore the non-conservation of matter.
An exchange of energy between the matter and the vacuum takes place, in such a way that  for $Q>0$ the vacuum decays into matter, whereas for $Q<0$ it is the other way around. Obviously we assume $0<|Q|\ll\dot{\rho}_m$ since we know that the standard conservation laws (\ref{standardlawmatter}) and (\ref{standardlawradiation}) are to a great extent correct.\\

\item \textbf{Type-G model}: $\dot{G}\neq 0$ {and} $\dot{\rho}_{\CC}\neq0$ assuming self-conservation of matter.

Since we assume the standard local covariant conservation laws
     in the separated forms (\ref{standardlawmatter}) and (\ref{standardlawradiation}), then Eq.\,(\ref{BianchiGeneral})
leads immediately to a dynamical interplay between the gravitational coupling and the vacuum energy density:
\begin{equation}\label{Bianchi1}
\dot{G}\,(\rmr+\rho_r+\rL)+G\dot{\rL}=0\,.
\end{equation} \\

\item \textbf{Type-AG model at fixed $\rL$}: $\dot{G}\neq 0$ and $\rL=$const.
Obviously this situation corresponds to non self-conservation of matter, and the matter densities obey
\begin{equation}\label{dGneqo}
\dot{G}(\rho_m+\rho_r+\rL)+ G\,\sum_{i=m,r}[\dot{\rho}_i+3H(\rho_i+p_i)]=0\,.
\end{equation}
Here we admit the possibility that $G$ does not stay constant with the cosmic evolution, but we assume that the $\CC$-term is a true cosmological constant. %
This situation is kind of complementary to the previous one since in the present instance there is a
dynamical interplay between $G$ and the anomalous matter conservation at fixed vacuum density.  A simpler and milder assumption would be to assume that radiation is conserved and that there is only exchange of energy between non-relativistic matter and vacuum. In this case it would simplify as follows:
\begin{equation}\label{dGneqo2}
\dot{G}(\rho_m+\rho_r+\rL)+G\,(\dot{\rho}_m+3H\rho_m)=0\,,
\end{equation}
where $\rho_r$ here is given by Eq,\,(\ref{standardlawradiation}).
The simplified form (\ref{dGneqo2}) could be solved e.g. for $G$,
if $\rmr$  would be given by some non-conservation
ansatz, or vice versa.

\item \textbf{General Type-AG model}: $\dot{G}\neq 0$, $\dot{\rho}_{\CC}=$const. \emph{and} anomalous matter conservation.
Needless to say this is the most general situation and is represented by Eq.\,(\ref{BianchiGeneral}), in which one may have both $G$ and $\rL$ dynamical, together with an anomalous matter conservation law. We may rewrite (\ref{BianchiGeneral}) in the more expanded form
\begin{equation}\label{type AG general}
\dot{G}(\rho_m+\rho_r+\rL)+ G\,\sum_{i=m,r}[\dot{\rho}_i+3H(\rho_i+p_i)]=Q=-\dot{\rho}_{\CC}\,.
\end{equation}
Obviously, Eq.\,(\ref{dGneqo}) is the particular case when $Q=0$.

\item \textbf{Inflationary Universe}: Another possibility with $\dot{G}\neq 0$ {and}
    $\dot{\rho}_{\CC}\neq 0$ is the case that there is no matter in
    the universe: $\rmr=\rho_r=0$. Then Eq.\,(\ref{BianchiGeneral}) obviously implies
    $G\,\rL=$const. Of course one possibility is that both $G$ and $\rL$ stay constant, as in the usual inflationary scenarios. But we cannot exclude that both parameters can be
    time evolving while the product remains constant. Whether $G$ and $\rL$ are both constant or not, such situation
    could only be of interest in the early universe, when matter still did not exist and only the vacuum energy was present.

\end{itemize}

 We will consider in the next sections some uses of the generalized cosmological scenarios indicated above. Although we will not address the type-AG models in this study, we will deal with some detail the type-A and type-G ones, as well as some more specific cases within the type-A models in which baryons and radiation are conserved but dark matter (DM) is not. The type-AG models are dealt with e.g. in\, \cite{FritzschSola2012,FritzschSola2015,FritzschSolaNunes2016}.

\section{The running vacuum model (RVM) and related models}

Among the general class of the DVM's, a particular subclass is the so-called running vacuum model (RVM), which can be motivated in the context of QFT in curved space-time (cf. \,\cite{JSPReview2013,SolaGomez2015} and references therein). In the RVM the dynamical nature of the vacuum is governed by a renormalization group equation. The model can be extended to provide an effective description of the cosmic evolution starting from the early inflationary phase of the universe\,\cite{JSPReview2013,SolaGomez2015,GRF2015,LimBasSol,Perico2013}. We will not consider here the applications to the early Universe since we want to test the model using the current data. We only need to know that well after inflation and up to our days, the RVM energy density can be written in the relatively simple form:
\begin{equation}\label{eq:RVMvacuumdadensity}
\rho_\Lambda(H) = \frac{3}{8\pi{G}}\left(c_{0} + \nu{H^2}\right)\,,
\end{equation}
where we are omitting the higher order terms ${\cal O}(H^4)$ which could be important only for the (inflationary) physics of the early universe.
The additive constant $c_0=H_0^2\left(\Omega_\Lambda-\nu\right)$ is fixed by the boundary condition $\rL(H_0)=\rLo$, where $\rLo$ and $H_0$ are the current values of these quantities, and $\OL$ is the normalized vacuum energy density with respect to the critical density now. The dimensionless coefficient $\nu$ encodes the dynamics of the vacuum and can be related with the $\beta$-function of the running. Thus, we naturally expect $|\nu|\ll1$. An estimate of $\nu$ in QFT indicates that it is of order $10^{-3}$ at most \cite{Fossil07}, but here we will treat $\nu$ as a free parameter and hence we shall deal with the RVM on pure phenomenological grounds. It means we will fit $\nu$ to the observational data.

A more general form of (\ref{eq:RVMvacuumdadensity}) for the current universe is when we include both the $H^2$ \emph{and} the $\dot{H}$ term, both being dimensionally consistent. In this case the vacuum energy density reads
\begin{eqnarray}
\rL(H;\nu,\alpha)=\frac{3}{8\pi G}\left(c_0+\nu H^2+\frac{2}{3}\alpha\,\dH\right)\,.\label{eq:A2}
\end{eqnarray}
Here, apart from $\nu$, there is also the second dimensionless coefficient, $\alpha$; altogether they parametrize the dynamics of vacuum. These coefficients can be related
with the $\beta$-function of the renormalization group running of the vacuum energy and therefore we naturally expect $|\nu|\ll1$ and $|\alpha|\ll1$.  In this generalized form of the RVM, the constant $c_0$ reads a bit more complicated: $c_0=H_0^2\left[\OLo-\nu+\alpha\left(\Omo+\frac43\,\Oro\right)\right]$.

In all formulations of the RVM we understand that $G$ can be either constant or variable with the cosmic evolution, in accordance to the various generic scenarios described in the previous section. We will next consider the RVM and other related models under some of these scenarios.

\subsection{Type-A  RVM's}

Let us display the explicit solution of model (\ref{eq:A2}) as a Type-A model, i.e when we have anomalous matter conservation at $G=$const.  One starts from Eq.\,\eqref{mixed conslaw}
\begin{equation}\label{eq:GeneralCL}
\dot{\rho}_r + 4H\rho_r + \dot{\rho}_m + 3H\rho_m = Q=-\dot{\rho}_\CC\,,
\end{equation}
with $\rL$ given in (\ref{eq:A2}).
After a straightforward calculation, in which we make use of $\dot{H}=-4\pi\,G\left[\rmr+ (4/3)\,\rho_r\right]$ and its time derivative $\ddot{H}=-4\pi\,G\left[\dot{\rho}_m+ (4/3)\,\dot\rho_r\right]\,$ so as to compute $\dot{\rho}_{\CC}$ from (\ref{eq:A2}),  it is easy to see that Eq.\,(\ref{eq:GeneralCL}) can be rewritten as follows:
\be\label{BianchitypeA2} {\dot \rho}_m +
3\,H\,\xi\,\rmr+\dot{\rho}_r+4\,H\,\xi'\,\rho_r=0\,,
\ee
where we have defined
\begin{equation}\label{defxixip}
\xi=\frac{1-\nu}{1-\alpha}\equiv 1-\nueff\,,\ \ \  \xi^\prime=\frac{1-\nu}{1-\frac{4}{3}\alpha}\equiv 1-\nueffp\,.
\end{equation}
For small $|\nu,\alpha|\ll1$ (the expected situation), we can write in good approximation $\nueff\simeq \nu-\alpha$ and $\nueffp\simeq \nu-(4/3)\alpha\ $ for these effective vacuum parameters.

The solution is more easily obtained in terms of the scale factor. Thus, trading the cosmic time for the scale factor $a$, i.e. using $d/dt=aH\,d/da$, we can easily solve (\ref{BianchitypeA2}) separately for the matter and radiation parts (the same kind of assumption as in the standard model case) as a function of $a$ . It immediately leads to the desired anomalous conservation laws for matter and radiation:
\begin{eqnarray}
\rho_m &=& \rho_m^0 ~a^{-3 \xi} \label{splitAsolutionM} \\
\rho_r &=& \rho_r^0 ~a^{-4 \xiR}\,.\label{splitAsolutionR}
\end{eqnarray}
The anomalies are of course related to the fact that in general we have $\xi\neq1$ and $\xi'\neq 1$ for $\nu,\alpha\neq0$. Since the matter and radiation densities are now known, the corresponding vacuum energy density follows from integrating (\ref{eq:GeneralCL}) once more in terms of the scale factor variable, with the result
\begin{equation}\label{rLArad}
\rL(a)=\rLo+{\rmo}\,\,(\xi^{-1} - 1) \left( a^{-3\xi} -1  \right) +
{\rRo}\,\,(\xi'^{-1} - 1) \left( a^{-4\xi'} -1\right)\,.
\end{equation}
The normalized  Hubble rate with respect to the current value, $E=H/H_0$, is easily obtained from Friedmann's equation, and we find:
\begin{equation}\label{HArad}
 E^2(a) =1+\frac{\Omo}{\xi}\left(a^{-3\xi}-1\right)+\frac{\ORo}{\xi'}\left(a^{-4\xi'}-1\right)\,.
\end{equation}
The normalized cosmological parameters with respect to the critical density satisfy the usual cosmic sum rule
\begin{equation}\label{sumruleRad}
 \Omo + \Oro+\OLo = 1\,.\end{equation}
 Needless to say the solution of the simpler model (\ref{eq:RVMvacuumdadensity}) ensues by setting $\alpha=0$ in the previous expressions, which leads to $\xi=\xi'=1-\nu$, and as a result for that model the anomaly in the radiation conservation law (\ref{splitAsolutionR}) coincides with the one in the (nonrelativistic) matter conservation law (\ref{splitAsolutionM}).  Finally we note that all of the above formulas immediately boil down to the $\CC$CDM case when $\nu=\alpha=0$ (i.e. $\xi=\xi'=1$), as they should.

 As an illustration, in Fig.\,1 we display the evolution of the vacuum energy density for the simpler model (\ref{eq:RVMvacuumdadensity}) as a function of the redshift $z=a^{-1}-1$. Specifically, we plot the relative variation with respect to the current value of the vacuum energy density:
 \begin{equation}
\label{VacuumEvolution}
 \frac{\Delta\rL}{\rLo}\equiv\frac{\rL(z)-\rLo}{\rLo} = \frac{\nu}{1 -\nu}\, \frac{\Omo}{\OLo}\, [(1+z)^{3(1-\nu)} -1]\,.
\end{equation}
Here we have neglected the radiation component since we are considering low values of $z$. For the plot in Fig.\,1 we use $\nu=10^{-3}$, which is the typical value obtained in the numerical fits discussed in  section 4, and is also the typical theoretical expectation for $\nu$ in QFT in curved spacetime\,\cite{Fossil07}. The corresponding anomalous (nonrelativistic) matter conservation law is\,\footnote{This equation was first proposed and tested in the literature in Ref.\,\cite{Cristina2003}, and only later on it was also exploited by other authors -- see e.g.\,\cite{WangMeng2004}.}
\begin{equation}
\label{MatterEvolution}
  \rho_m(z) = \rho_{m}^0 (1+z)^{3(1- \nu)}\,.
\end{equation}
By inserting $\rL(z)$ and $\rmr(z)$, from (\ref{VacuumEvolution}) and (\ref{MatterEvolution}) respectively, in the generalized conservation law (\ref{mixed conslaw}) for type-A models (in the matter-dominated epoch)  we find that it is identically satisfied, as it should.

\begin{figure}[t]
\begin{center}
\includegraphics[width=3in, height=3in]{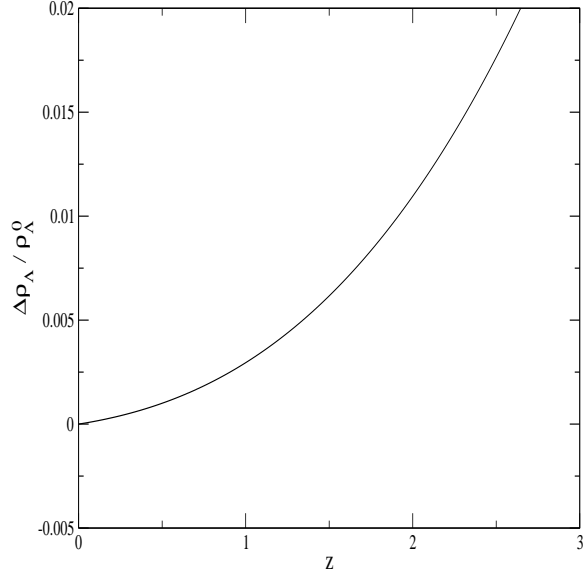}
\caption{The evolution of the vacuum energy density within the RVM for $\nu=0.001$ We plot the relative variation of $\rL(z)$ with respect to the current value, see Eq.\,(\ref{VacuumEvolution}).}
\end{center}
\label{G2}
\end{figure}


\subsection{RVMc: 
the running vacuum model with conserved baryon and radiation densities}

In this section we still consider type-A RVM's but now introduce a further specification about the conservation of matter which was not imposed thus far. Recall that $\rho_m$ can be split into the contribution from baryons and cold dark matter (DM), namely $\rho_m=\rho_b+\rho_{dm}$. In the following we assume that the dark matter density is the only one that carries the anomaly, whereas radiation and baryons are self-conserved, so that their energy densities evolve in the standard way (\ref{standardlawmatter}-\ref{standardlawradiation}), i.e. $\rho_r(a)=\rho_{r}^0\,a^{-4}$ and $\rho_b(a) = \rho_{b}^0\,a^{-3}$. We call this version of the RVM, the RVMc. In this case we limit ourselves to consider the case when $\alpha=0$ in (\ref{eq:A2}). In other words, we focus now on the canonical RVM (\ref{eq:RVMvacuumdadensity}). The solution is not just (\ref{splitAsolutionM}-\ref{rLArad}) with $\xi=\xi'=1-\nu$ since Eq.\,(\ref{eq:GeneralCL}) cannot be written in the form (\ref{BianchitypeA2}) for the entire matter density $\rho_m=\rho_b+\rho_{dm}$ owing to the conservation of the individual $\rho_b$ and $\rho_r$ components. Imposing this conservation condition in Eq.\,(\ref{eq:GeneralCL})  we find that the interaction of matter with the vacuum becomes exclusively associated to the exchange of energy with the dark sector, and hence we can rewrite (\ref{eq:GeneralCL}) as follows:
\begin{equation}\label{eq:Qequations}
\dot{\rho}_{dm}+3H\rho_{dm}=Q\,,\ \ \ \ \ \ \, \dot\rho_{\CC}=-{Q}\,.
\end{equation}
It is for this reason that the conservation law for the DM component is anomalous. The source function $Q$ is a calculable expression from \eqref{eq:RVMvacuumdadensity} and Friedmann's equation. We find:
\begin{equation}\label{eq:QRVM}
{\rm RVMc:}\qquad Q=-\dot{\rho}_{\Lambda}=\nu\,H(3\rho_{dm}+3\rho_{b}+4\rho_r)\,.
\end{equation}
Finally, we can solve routinely for the energy densities of the DM and vacuum and we arrive at the following formulas:
\begin{eqnarray}\label{eq:rhomRVMc}
\rho_{dm}(a) &=& \rho_{dm}^0\,a^{-3(1-\nu)} + \rho_{b}^0\left(a^{-3(1-\nu)} - a^{-3}\right)- \frac{4\nu\rho^{0}_r}{1+3\nu}\left(a^{-4}-a^{-3(1-\nu)}\right)
\end{eqnarray}
and
\begin{eqnarray}\label{eq:rLRVMc}
\rho_\Lambda(a) &=& \rLo + \frac{\nu\,\rho_{m}^0}{1-\nu}\left(a^{-3(1-\nu)}-1\right)\nonumber\\
 &+& \frac{\nu}{1-\nu}\,\,\rho^{0}_{r}\left(\frac{1-\nu}{1+3\nu}\,a^{-4} + \frac{4\nu}{1+3\nu}\,a^{-3(1-\nu)}-1\right).
\end{eqnarray}
As can be easily checked, for $\nu\to 0$ we recover the corresponding results for the $\CC$CDM, as we should.
The Hubble function can be immediately obtained from these formulas after inserting them in Friedmann's equation, together with the conservation laws for baryons and radiation, $\rho_r(a)=\rho_{r}^0\,a^{-4}$ and $\rho_b(a) = \rho_{b}^0\,a^{-3}$. We refrain from writing out these expressions.

\subsection{Other DVM's with conserved baryon and radiation densities}

Let us now explore two additional type-A phenomenological DVM's with conserved baryon and radiation densities, in which the source function $Q$ in (\ref{eq:Qequations}) is introduced purely \textit{ad hoc}, i.e. without any special theoretical motivation. Recall that in the RVMc case discussed previously we could calculate $Q$ because $\rL$ was known from Eq.\,(\ref{eq:RVMvacuumdadensity}). Now we leave this RVMc scenario for a while and  introduce alternative forms for $Q$ on merely phenomenological grounds.  Two possible ansatzs discussed in the literature are the following:
\begin{eqnarray}\label{eq:PhenModelQdm}
{\rm Model\ \ }Q_{dm}: \phantom{XX}Q&=&3\nu_{dm}H\rho_{dm}\\
{\rm Model\ \ }Q_{\CC}:\phantom{XXx}Q&=&3\nu_{\Lambda}H\rho_{\Lambda}\,.\label{eq:PhenModelQL}
\end{eqnarray}
Model $Q_{\Lambda}$  was previously studied e.g. in \cite{Salvatelli2014}, but as mentioned in \cite{Comparison2016,PRDcompanion} we do not concur with their analysis, see also \,\cite{Murgia2016}. Model $Q_{dm}$ was considered recently in\,\cite{Li2016}. It is closer to the RVM than $Q_{\Lambda}$, but not identical, compare equations\,(\ref{eq:QRVM}) and (\ref{eq:PhenModelQdm}).

In the above alternative models we have introduced new dimensionless coefficients $\nu_{dm}$ and $\nu_{\Lambda}$, which are in principle differen from $\nu$ for the RVM.  Altogether these coefficients $\nu_{i}=(\nu,\nu_{dm},\nu_\Lambda)$  parametrize the evolution of the vacuum energy density and the strength of the dark-sector interaction for each model. For $\nu_{i}>0$ (hence $Q>0$) the vacuum decays into dark matter (which is favorable from the point of view of the second law of thermodynamics) whereas for $\nu_{i}<0$ ($Q<0$) is the other way around.
We can easily account for models (\ref{eq:PhenModelQdm}) and (\ref{eq:PhenModelQL}) by solving the differential equations (\ref{eq:Qequations}) in each case.

As for model $Q_{dm}$ we find:
\begin{eqnarray}
\rho_{dm}(a) &=& \rho_{dm}^0\,a^{-3(1-\nu_{dm})}
\label{eq:rhoms2} \\
\rho_\Lambda(a)&=& \rLo + \frac{\nu_{dm}\,\rho_{dm}^0}{1-\nu_{dm}}\,\left(a^{-3(1-\nu_{dm})}-1\right)\,,
\end{eqnarray}
whereas for model $Q_{\Lambda}$:
\begin{eqnarray}
\rho_{dm}(a) &=&\rho_{dm}^0\,a^{-3} + \frac{\nu_\Lambda}{1-\nu_\CC}\rLo\left(a^{-3\nu_\Lambda}-a^{-3}\right)\nonumber\\
\rho_\Lambda(a) &=&\rLo\,{a^{-3\nu_\Lambda}}\,.
\end{eqnarray}
In section 4 we will compare these alternative DVM's with the RVM in their different ability to describe the cosmological data, and of course we will compare them all to the $\CC$CDM to check which model performs better observationally.

\subsection{Type-G  RVM's}

Let us consider anew the more general sort of RVM indicated in (\ref{eq:A2}). However, we shall deal with it here as a type-G model, meaning that $G$ is now a cosmological variable that satisfies the differential equation (\ref{Bianchi1}) with both conservation of nonrelativistic matter and radiation separately. Trading once more the cosmic time for the scale factor $a$
in that equation we can integrate it and determine $G$ as a function of $a$. This can be done as follows. Combining Friedmann's equation (\ref{eq:FriedmannEq}) and the acceleration equation (\ref{eq:PressureEq}), and using the matter conservation equations, we arrive at
\be\label{eq:Star} G(a)=- G_0\,\left[\frac{a\,dE^2(a)/da}{3 \Omo\,a^{-3}+4\ORo\,a^{-4}}\right]\,, \ee
where $G_0\equiv G(a=1)$ is the present value of $G$, and $E(a)=H(a)/H_0$ is the normalized Hubble rate to its present value.  The above equation links $G(a)$ to $dE^2(a)/da$. The final step is to insert (\ref{eq:Star}) and (\ref{eq:A2}) into Eq.\,(\ref{eq:FriedmannEq}), and integrating the resulting differential equation for $E^2(a)$:
\begin{equation}\label{eq:E2a}
\frac{\frac{a}{\xi}+ \frac{\Omega^0_r}{\xi'\Omega^0_m}} {3+4\frac{\Omega^0_r}{\Omega^0_m}a^{-1}}\,\frac{dE^2}{da}+E^2 - \frac{\OLo-\nu+\alpha\left(\Omo+\frac43\,\Oro\right)}{1-\nu} = 0 \,.
\end{equation}
The result after some calculations is:
\begin{eqnarray}\label{eq:DifEqH} &&E^2(a)=1+\left(\frac{\Omega_m^0}{\xi}+\frac{\Omega_r^0}{\xi^\prime}\right)
\times\left[-1+a^{-4\xi^\prime}\left(\frac{a\xi^\prime+\xi\Omega_r^0/\Omega_m^0}{\xi^\prime+\xi\Omega_r^0/\Omega_m^0}\right)^{\frac{\xi^\prime}{1-\alpha}}\right]\,, \end{eqnarray}
where $\xi$ and $\xi'$ have been defined previously in (\ref{defxixip}).
In the radiation-dominated epoch, the leading behavior of Eq.\,(\ref{eq:DifEqH}) is $\sim \ORo\,a^{-4\xi'}$, while in the matter-dominated epoch is $\sim \Omo\,a^{-3\xi}$. Furthermore, for $\nu,\alpha\to 0$,  one can easily show that $E^2(a)\to 1+\Omo\,(a^{-3}-1)+\ORo(a^{-4}-1)$. This is the $\CC$CDM form, as expected in that limit.  From the above rather unwieldy expression for $E(a)$ one can finally obtain the explicit form of $G(a)$ by computing the derivative $dE^2(a)/da$ and inserting it in (\ref{eq:Star}).  The explicit scale factor dependence of the vacuum energy density, i.e. $\rL=\rL(a)$,
ensues similarly upon inserting  \eqref{eq:DifEqH} into \eqref{eq:A2}.
We refrain from writing out these cumbersome expressions and we limit ourselves to quote some simplified forms. For instance, the form of $\rL(a)$ when we can neglect the radiation contribution is simple enough to be quoted here:
\be\label{eq:RhoLNR} \rL(a)=\rho_{c}^0\,
a^{-3}\left[a^{3\xi}+\frac{\Omo}{\xi}(1-\xi-a^{3\xi})\right]\,.
\ee
\begin{table}
\begin{center}
\caption{\scriptsize
The best-fit values for the $\CC$CDM, XCDM and the RVM's, including their statistical  significance ($\chi^2$-test and Akaike and Bayesian information criteria, AIC and BIC, see Sect. \ref{sect:AkaikeBayes}). The large and positive values of $\Delta$AIC and $\Delta$BIC strongly favor the dynamical DE options (RVM's and XCDM) against the $\CC$CDM (see text). We use a total of $89$ data points from SNIa+BAO+$H(z)$+LSS+CMB observables in our fit: namely $31$ points from the JLA sample of SNIa\,\cite{SNmod}, $11$ from BAO\,\cite{Beutler2011,Ross2015,Kazin2014,GilMarin2016,Delubac2015,Aubourg2015}, $30$  from $H(z)$\,\cite{Zhang2014,Jimenez2003,Simon2005,Moresco2012,Moresco2016,Stern2010,Moresco2015}, $13$ from linear growth \cite{GilMarin2016,Beutler2012,Feix2015,Simpson2016,Blake2013,Blake2011BAO,Springob2016,Granett2015,Guzzo2008,SongandPercival2009}, and $4$ from CMB\,\cite{Huang2015}. For a summarized description of these data, see \cite{FirstEvidence2016}. In the XCDM model the EoS parameter $\omega$ is left free, whereas for the RVM's and $\CC$CDM is fixed at $-1$.  The specific RVM fitting parameter is $\nueff$, see Eq.\,(\ref{defxixip}) and the text. For G1 and A1 models, $\nueff=\nu$. The remaining parameters are the standard ones ($h,\omega_b,n_s,\Omega_m$).  The quoted number of degrees of freedom ($dof$) is equal to the number of data points minus the number of independent fitting parameters ($5$ for the $\CC$CDM, $6$ for the RVM's and the XCDM).  The normalization parameter M introduced in the SNIa sector of the analysis is also left free in the fit,  but it is not listed in the table. For the CMB data we have used the marginalized mean values and standard deviation for the parameters of the compressed likelihood for Planck 2015 TT,TE,EE + lowP data from \cite{Huang2015}, which provide tighter constraints to the CMB distance priors than those presented in \cite{PlanckDE2015}.}
\label{Table1}
\begin{scriptsize}
\resizebox{1\textwidth}{!}{
\begin{tabular}{| c | c |c | c | c | c | c | c | c | c |}
\multicolumn{1}{c}{Model} &  \multicolumn{1}{c}{$h$} &  \multicolumn{1}{c}{$\omega_b= \Omega_b h^2$} & \multicolumn{1}{c}{{\small$n_s$}}  &  \multicolumn{1}{c}{$\Omega_m$}&  \multicolumn{1}{c}{{\small$\nu_{eff}$}}  & \multicolumn{1}{c}{$\omega$}  &
\multicolumn{1}{c}{$\chi^2_{\rm min}/dof$} & \multicolumn{1}{c}{$\Delta{\rm AIC}$} & \multicolumn{1}{c}{$\Delta{\rm BIC}$}\vspace{0.5mm}
\\\hline
$\Lambda$CDM  & $0.693\pm 0.003$ & $0.02255\pm 0.00013$ &$0.976\pm 0.003$& $0.294\pm 0.004$ & - & $-1$  & 90.44/85 & - & - \\
\hline
XCDM  & $0.670\pm 0.007$& $0.02264\pm0.00014 $&$0.977\pm0.004$& $0.312\pm0.007$ & - &$-0.916\pm0.021$  & 74.91/84 & 13.23 & 11.03 \\
\hline
A1  & $0.670\pm 0.006$& $0.02237\pm0.00014 $&$0.967\pm0.004$& $0.302\pm0.005$ &$0.00110\pm 0.00026 $ &  $-1$ & 71.22/84 & 16.92 & 14.72 \\
\hline
A2   & $0.674\pm 0.005$& $0.02232\pm0.00014 $&$0.965\pm0.004$& $0.303\pm0.005$ &$0.00150\pm 0.00035 $& $-1$  & 70.27/84 & 17.87 & 15.67\\
\hline
G1 & $0.670\pm 0.006$& $0.02236\pm0.00014 $&$0.967\pm0.004$& $0.302\pm0.005$ &$0.00114\pm 0.00027 $& $-1$  &  71.19/84 & 16.95 & 14.75\\
\hline
G2  & $0.670\pm 0.006$& $0.02234\pm0.00014 $&$0.966\pm0.004$& $0.303\pm0.005$ &$0.00136\pm 0.00032 $& $-1$  &  70.68/84 & 17.46 & 15.26\\
\hline
\end{tabular}}
 \end{scriptsize}
\end{center}
\end{table}
%
In the expression above, $\rho_{c}^0=3H_0^2/8\pi\,G_0$ is the current critical density and $G_0\equiv G(a=1)$ is the current value of the gravitational coupling. Quite obviously, for $\xi=1$ we recover the $\CC$CDM form: $\rL=\rho_{c}^0(1-\Omo)=\rho_{c}^0\OLo=$const. As for the gravitational coupling, it
evolves logarithmically with the scale factor in the limit $\nu,\alpha\to 0$ and hence changes very slowly\footnote{This is a welcome feature already expected in particular  realizations of type-G models in QFT in curved spacetime\,\cite{Fossil07}.}. As mentioned, the exact form of $G(a)$ is a bit too bulky. It suffices to say here that it behaves as
\begin{equation}\label{Gafunction}
G(a)=G_0\,a^{4(1-\xi')}\,f(a)\simeq G_{0}(1+4\nueffp\,\ln\,a)\,f(a)\,,
\end{equation}
where $f(a)=f(a;\Omo,\ORo; \nu,\alpha)$ is a smooth function of the scale factor, which tends to one at present ($f(a)\to 1$ for $a\to 1$) irrespective of the values of the various parameters $\Omo,\ORo,\nu,\alpha$ involved in it; and $f(a)\to1$ also in the remote past ($a\to 0$) for $\nu,\alpha\to 0$ (i.e. $\xi,\xi'\to 1$). As expected, $G(a)\to G_0$ for $a\to 1$, and $G(a)$ has a logarithmic evolution for $0<|\nueff|\ll 1$.
Notice that the limit $a\to 0$ is  relevant for the BBN (Big Bang Nucleosynthesis) epoch and therefore $G(a)$ should not depart too much from $G_0$ according to the usual bounds on BBN. This restriction is tacitly incorporated in our numerical analysis of the RVM models in the next section.

%
\begin{figure*}
\centering
\includegraphics[angle=0,width=0.9\linewidth]{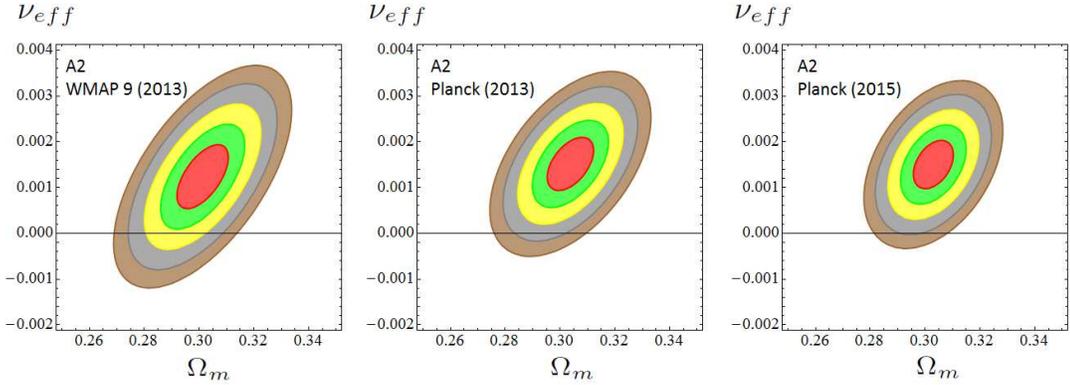}
\caption{\label{fig:A2Evolution}%
\scriptsize Likelihood contours in the $(\Omega_m,\nueff)$ plane for the values $-2\ln\mathcal{L}/\mathcal{L}_{max}=2.30$, $6.18, 11.81$, $19.33$, $27.65$ (corresponding to 1$\sigma$, 2$\sigma$, 3$\sigma$, 4$\sigma$ and 5$\sigma$ c.l.) after marginalizing over the rest of the fitting parameters indicated in Table 1. We display the progression of the contour plots obtained for model A2 using the 90 data points on SNIa+BAO+$H(z)$+LSS+BBN+CMB, as we evolve from the high precision CMB data from WMAP9\,\cite{WMAP9-2013}, Planck 2013\,\cite{PLANCK2013} and Planck 2015\,\cite{PLANCK2013}, see \,\cite{FirstEvidence2016} for details. In the sequence, the prediction of the concordance  model ($\nueff=0$) appears increasingly more disfavored, at an exclusion c.l. that ranges from  $\sim 2\sigma$ (for WMAP9), $\sim 3.5\sigma$ (for Planck 2013) and up to  $4\sigma$ (for Planck 2015). Using the Fisher matrix and numerical integration, we find that $\sim99.82\%$ of the area of the $4\sigma$ contour for Planck 2015 (and  $\sim95.49\%$ of the corresponding $5\sigma$ region) satisfies $\nueff>0$.  The $\CC$CDM  becomes once more excluded at $\sim 4\sigma$ c.l. Subsequent marginalization over $\Omega_m$ increases slightly the c.l. and renders the fitting values indicated in Table 1, which reach a statistical significance of $4.2\sigma$ for all the RVM's. The corresponding AIC and BIC criteria (cf. Table 1)
consistently imply a very strong support to the RVM's against the $\CC$CDM.
}
\end{figure*}

\section{Fitting the DVM's to observations}

In this section, we put the vacuum models discussed above to the test, see \cite{ApJ2015,JCAP2015b,FirstEvidence2016, Comparison2016,PRDcompanion,MNRAS2015,JCAP2015a,BPS2009,Grande2011} for more details. We consider the general RVM model (\ref{eq:A2}), which depends on the two vacuum parameters $\nu$ and $\alpha$, under the two modalities type-A and type-G. Let us start with  type-A (cf. section 3.1)  and then we will address the type-G case (cf. section 3.4). Subsequently we will focus on the RVMc in the canonical form (\ref{eq:RVMvacuumdadensity}), together with the $Q_m$ and $Q_{\CC}$ models (cf. sections 3.2 and 3.3). It proves useful to study them altogether in a comparative way.

\subsection{Type-A and type-G}

We confront now the various DVM's to the main set of cosmological observations compiled to date, namely we fit the models to the following wealth of data: i) the data from distant type Ia supernovae (SNIa); ii) the data on baryonic acoustic oscillations (BAO's); iii) the known values of the Hubble parameter at different redshift points, $H(z_i)$; iv) the large scale structure (LSS) formation data encoded in the weighted linear growth rate $f(z_i)\sigma_8(z_i)$; v) the BBN bound on the Hubble rate (when applicable); and, finally, vi) the CMB distance priors from WMAP9 and Planck 2013 and 2015.  In short, we use the essential observational data represented by the cosmological observables SNIa+BAO+$H(z)$+LSS+BBN+CMB to fit the vacuum parameters\,\footnote{The bounds on the vacuum parameters from the analysis of cosmic perturbations, including the power spectrum, are also compatible with the fitting
results presented here, cf.\,\cite{PericoTamayo}. See also \cite{AdriaJoan2017} for a detailed analysis of the matter and vacuum perturbations in running vacuum models.}.  For the analysis we have defined a joint likelihood function ${\cal L}$ from the product of the likelihoods for all the data sets discussed above. For Gaussian errors, the total $\chi^2$ to be minimized reads:
\be
\chi^2_{tot}=\chi^2_{SNIa}+\chi^2_{BAO}+\chi^2_{H}+\chi^2_{f\sigma_8}+\chi^2_{CMB}\,.
\ee
Each one of these terms is defined in the standard way,
including the covariance matrices for each sector\,\cite{AmendolaStatistics}.
The fitting results are given in Table 1, see\,\cite{FirstEvidence2016,PRDcompanion} for more details. The contour plots  for type-A and type-G are displayed  in Figs.\,2 and 3, respectively. In these figures the models referred to as A2 and G2 mean that we are fitting the two parameters $\nu$ and $\alpha$ of (\ref{eq:A2}). In the plots we indicate the result for $\nueff=\nu-\alpha$, which we defined previously in (\ref{defxixip}). We can see in these figures the increasingly favorable evolution of the fits in support of these DVM's when we successively use data from WMAP9\,\cite{WMAP9-2013}, Planck 2013\,\cite{PLANCK2013} and Planck 2015\,\cite{PLANCK2013}. This situation is not particular of A2 and G2, Table 1 clearly shows that if we fit only the parameter $\nu$ (i.e. taking $\alpha=0$, corresponding to models A1 and G1), we find always a similar quality fit\,\cite{FirstEvidence2016}. In all these cases, remarkably enough, the fit is significantly better than the $\CC$CDM one. We can see in Figs.\,2 and 3 that we can indeed reach $4\sigma$ evidence that at least one of these vacuum parameters is nonvanishing and positive, i.e. $\sim 4\sigma$ evidence against the $\CC$CDM. The fitting results indicated in Table 1 actually lead to $4.2\sigma$ evidence owing to the fact that in the final numerical results we have marginalized over $\Omo$ as well. Therefore, the c.l. that we can read from the fitting tables is always slightly higher than the one that can be visually observed from the contour plots.

\begin{figure*}
\centering
\includegraphics[angle=0,width=0.9\linewidth]{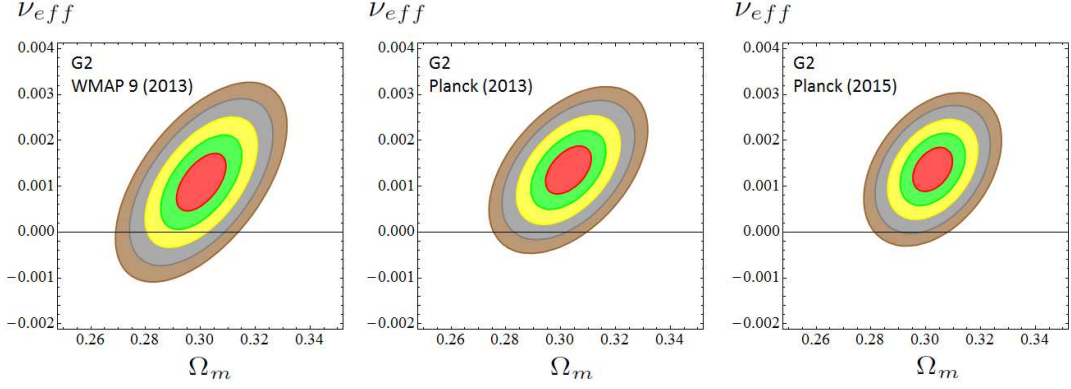}
\caption{\label{fig:A1Evolution}%
\scriptsize As in Fig.\,1, but for model G2. Again we see that the contours tend to migrate to the $\nueff>0$ half plane as we evolve from WMAP9\,\cite{WMAP9-2013} to Planck 2013\,\cite{PLANCK2013} and Planck 2015\,\cite{PLANCK2015} data. Using the same method as in Fig.\,1, we find that $\sim99.81\%$ of the area of the $4\sigma$ contour for Planck 2015 (and  $\sim95.47\%$ of the corresponding $5\sigma$ region) satisfies $\nueff>0$.  The $\CC$CDM  becomes once more excluded at $\sim 4\sigma$ c.l.
}
\end{figure*}

\subsection{Fiducial model}

We should clarify an important aspect of the fitting results that we  obtain for the dynamical vacuum models under study, namely the fact that all models are compared to the same fiducial model (including the $\Lambda$CDM). This is important in order to fix the normalization factor
of the power-spectrum. As a fiducial model we take the $\CC$CDM model in which all parameters are taken from the Planck 2015 TT,TE,EE+lowP+lensing analysis\, \cite{PLANCK2015}. Let us explain in more detail the procedure by considering our treatment of the linear structure formation data. We have computed the density contrast $\delta_m=\delta\rho_m/\rho_m$ for each vacuum model by adapting the cosmic perturbations formalism for type-A and type-G vacuum models.
The matter perturbation, $\delta_m$, obeys a generalized differential equation which depends on the RVM type. For type-A models such equation with respect to the cosmic time reads\,\footnote{For a derivation and more details on this equation, see the comprehensive works \cite{JCAP2015a}, \cite{MNRAS2015} and \cite{JCAP2015b}.}
\begin{equation}\label{diffeqD}
\ddot{\delta}_m+\left(2H+\Psi\right)\,\dot{\delta}_m-\left(4\pi
G\rmr-2H\Psi-\dot{\Psi}\right)\,\delta_m=0\,,
\end{equation}
where $\Psi\equiv-\frac{\dot{\rho}_{\CC}}{\rmr}$. For $\rL=$const. we have $\Psi=0$ and Eq.\,(\ref{diffeqD}) reduces, of course, to the $\CC$CDM form. For further convenience, let us write the last term of the $\CC$CDM perturbation equation in terms of the Hubble function:
\begin{equation}\label{eq:DifLCDM0}
\ddot{\delta}_m+2H\dot{\delta}_m+\dot{H}\delta_m=0\,.
\end{equation}
For type-G models the matter perturbation equation is more complicated \cite{ApJ2015,JCAP2015b}:
\be\label{eq:DifEqCosmicTime}
\dddot{\delta}_m+5H\ddot{\delta}_m+3\dot{\delta}_m(\dot{H}+2H^2)=0\,.
\ee
This is a third order equation, but it is straightforward to show that the time derivative of the ordinary second order equation for the perturbations, i.e. the derivative of Eq.\,(\ref{eq:DifLCDM0}), coincides with (\ref{eq:DifEqCosmicTime}) when $\CC=$const., as it should\,\cite{JCAP2015b}. However for dynamical $\CC$ the perturbation equation that holds for type-G models is Eq.\,(\ref{eq:DifEqCosmicTime}) -- see also \cite{Grande2011}.


From the above perturbation equations we can derive the weighted linear growth $f(z)\sigma_8(z)$ for any of the type-A or type-G models, where $f(z)=d\ln{\delta_m}/d\ln{a}$ is the growth factor and $\sigma_8(z)$ is the rms mass fluctuation amplitude on scales of $R_8=8\,h^{-1}$ Mpc at redshift $z$. The latter is computed from
\begin{equation}
\begin{small}\sigma_{\rm 8}(z)=\sigma_{8, \Lambda}
\frac{\delta_m(z)}{\delta^{\CC}_{m}(0)}
\left[\frac{\int_{0}^{\infty} k^{n_s+2} T^{2}(\vec{p},k)
W^2(kR_{8}) dk} {\int_{0}^{\infty} k^{n_{s,\CC}+2} T^{2}(\vec{p}_\Lambda,k) W^2(kR_{8,\Lambda}) dk}
\right]^{1/2}\label{s88general}
\end{small}\end{equation}
with $W$ a top-hat smoothing function (see e.g. \cite{JCAP2015a} for
a more elaborated discussion) and $T(\vec{p},k)$ the transfer function, which we take from\,\cite{Bardeen}. In addition, we have defined the fitting vectors $\vec{p}=(h,\omega_b,n_s,\Omega_m,\nueff)$ for the vacuum models we are analyzing (including the $\Lambda$CDM), and $\vec{p}_\CC=(h_{\CC},\omega_{b,\CC},n_{s,\CC},\Omega_{m,\CC},0)$ for the fiducial $\CC$CDM model that we use in order to fix the normalization factor of the power-spectrum.  From equation (\ref{s88general}) we can now understand the point under discussion. The denominator of this equation is fully defined in terms of the aforementioned fiducial model in which all cosmological parameters are taken from the Planck 2015 TT,TE,EE+lowP+lensing analysis, \cite{PLANCK2015}. These parameters are denoted by a subindex $\CC$. As we can see, the calculation of $\sigma_{\rm 8}(z)$ and therefore of the  the weighted linear growth $f(z)\sigma_8(z)$, is computed for all models by taking the fiducial model as a reference. The comparison of the theoretical predictions and the data are indicated in Fig. 4. Since the $\CC$CDM model is also compared to that fiducial model, all models ($\CC$CDM, RVM's and XCDM) are normalized in the same way. Clearly, this is an optimal strategy to compare the dynamical DE models to the $\CC$CDM in the fairest possible way\footnote{This point was treated in a different way in a previous version of our analysis and this led to larger values of the AIC and BIC parameters for the dynamical models. However, after normalizing all models with respect to the same fiducial model the fitting results remain essentially unaltered and the $\Delta$AIC and $\Delta$AIC differences in favor of the dynamical DE option remain rather large, namely large enough to be ranked in the ``strong'' to ``very strong'' range of evidence in favor of dynamical DE. This is clear from Table 1 and also from Table 2 below. }.

\begin{figure}
\centering
\includegraphics[angle=0,width=0.7\linewidth]{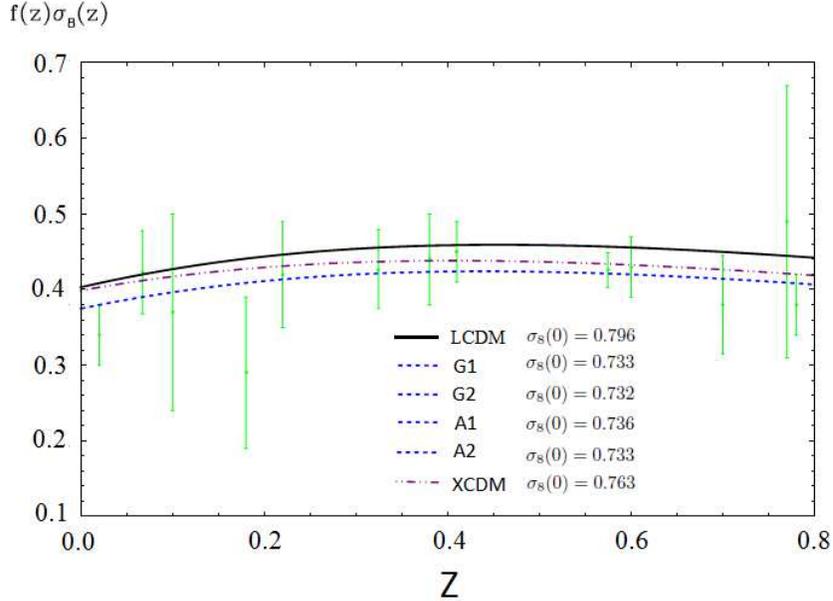}
\caption{\label{fig:fsigma8}%
\scriptsize {The $f(z)\sigma_8(z)$ data and the predicted curves by the RVM's, XCDM and the $\CC$CDM, using the best-fit values in Table 1. Shown are also the values of $\sigma_8(0)$ that we obtain for all the models. The theoretical prediction of all the RVM's are visually indistinguishable and they have been plotted using the same (blue) dashed curve. The observational data points used (in green) and corresponding observational references are given in \cite{FirstEvidence2016}, see Table 4 of that reference.}
}
\end{figure}
%

%
\subsection{Comparing with the XCDM model}

In the XCDM model one replaces the $\CC$-term with an unspecified dynamical entity $X$, whose energy density at present coincides with the current value of the vacuum energy density, i.e. $\rho_X^0=\rLo$. It is not even a model, it is however the simplest possible parametrization for the dynamical DE, in which the equation of state (EoS) parameter is taken as constant, i.e. $p_X=\omega\,\rho_X$ with $\omega=$conts. Such parametrization was first introduced long ago by Turner and White\,\cite{TurnerWhite1997}. In a sense it mimics the behavior of a scalar field, quintessence ($\omega\gtrsim-1$) or phantom ($\omega\lesssim-1$), under the assumption that such field has an essentially constant EoS parameter near (but not exactly) $-1$. Since both matter and DE are self-conserved (i.e., they are not interacting), the energy densities as a function of the scale factor are simply given by $\rho_m(a)=\rho_m^0\,a^{-3}$ and $\rho_X(a)=\rho_X^0\,a^{-3(1+\omega)}$.
The Hubble function is therefore given by
\begin{equation}\label{eq:HXCDM}
H^2(a)=\frac{8\pi G}{3}\left[\rho_m^0\,a^{-3}+\rho_X^0\,a^{-3(1+\omega)}\right]=H_0^2\left[\Omega_m^0\,a^{-3}+(1-\Omega_m^0)\,a^{-3(1+\omega)}\right]\,.
\end{equation}
\begin{figure*}[t]
\centering
\includegraphics[angle=0,width=0.40\linewidth]{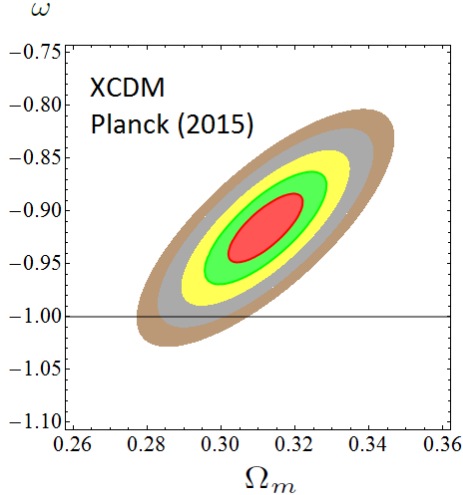}
\caption{\label{fig:XCDMEvolution}%
\scriptsize As in Fig.\,1 and 2, but for model XCDM. The $\CC$CDM is excluded at $\sim 4\sigma$ c.l.
}
\end{figure*}

A more sophisticated approximation to the behavior of a scalar field playing the role of dynamical DE is provided by the CPL prametrization\,\cite{CPL}, given by
\begin{equation}\label{eq:CPL}
\omega=\omega_0+\omega_1\,(1-a)=\omega_0+\omega_1\,\frac{z}{1+z}\,.
\end{equation}
However, it involves two parameters $(\omega_0,\omega_1)$. Therefore, in order to better compare with the one-parameter family of dynamical vacuum models under consideration it is better at this point to stick to the XCDM parametrization, which has also one parameter $\omega$ only.  Recall that the RVM's have the parameter $\nu$ (models G1,A1) or the two parameters $(\nu,\alpha)$ (models G2,A2), but in the last case we fix a convenient relation between $\nu$ and $\alpha$, so any of these models has actually one-parameter in the DE sector\,\cite{FirstEvidence2016}. For this reason the XCDM, with also one single parameter, is more appropriate for comparison with the RVM's. The XCDM serves as a baseline dynamical DE model to compare any more sophisticated model for the dynamical DE. If a given model claims to be sensitive to dynamical DE effects (as it is the case with the RVM's and other DVM's considered here), it is convenient to see if part or all of these effects can be detected through the XCDM parametrization. Let us note that we should not necessarily expect that the XCDM is sensitive to the DE effects in a way comparable to the DVM's. The latter, for example, are in some cases vacuum models that interact with matter or have a variable $G$. Nothing of this applies for the XCDM, so the mimicking of the vacuum dynamics need not be perfect through the XCDM. Still, since the vacuum dynamics is mild enough, we expect that a significant part of the vacuum dynamics should be captured by the XCDM parametrization, either in the form of effective quintessence behavior ($\omega\gtrsim -1$) or effective phantom behavior ($\omega\lesssim-1$).  The result can be clearly appraised in Tables 1 and 2 (where the best fit parameters for the XCDM are also quoted), as well as in the contour plots of Fig. 5. We see that the effective quintessence option is definitely projected at exactly $4\sigma$ level: $\omega=-0.916\pm0.021$. Thus, a large part of the dynamical vacuum effect suggested by the RVM's (at $4.2\sigma$ c.l.) is also disclosed  by the XCDM parametrization.

Recently, in Ref.\cite{phiCDM16} further evidence on the time-evolving nature of the dark energy has been provided by fitting the same cosmological data as in the current work in terms of specific scalar field models. As a representative model in that work it was used the original Peebles \& Ratra potential, $V\propto\phi^{-\alpha}$. Remarkably enough, unambiguous signs of dynamical DE at  $\sim 4\sigma$ c.l. have also been found, thus reconfirming through a nontrivial scalar field approach the strong hints of dynamical DE found with the dynamical vacuum models and the XCDM  parametrization.

\subsection{RVMc, $Q_m$ and $Q_{\CC}$ with conserved baryons and radiation}

In Fig.\,6, based on Table 2, we continue our numerical analysis by considering once more the fitting results to the the cosmological observables SNIa+BAO+$H(z)$+LSS+CMB, but we now address the triad of models RVMc and $Q_m$ and $Q_{\CC}$. Recall that for these particular models we assume that baryons and radiation are conserved and the interaction of the vacuum energy is only with dark matter  (cf. sections 3.2 and 3.3).  Thus, no BBN bound is necessary in this case since the radiation content at the BBN time is identical to that of the $\CC$CDM. Also for simplicity we set $\alpha=0$ in Eq.\,(\ref{eq:A2}), so that the vacuum energy density takes here the simpler form (\ref{eq:RVMvacuumdadensity}). The details on the SNIa+BAO+$H(z_i)$+LSS+CMB data points used are given in Ref.\,\cite{Comparison2016,PRDcompanion}. Here we limit ourselves to show the final fitting results, see Table 2. We emphasize the use of updated observational inputs concerning BAO and $f(z)\sigma_8(z)$ linear growth data, particularly from \cite{GilMarin2016} -- rather than the older data from \cite{GilMarin2015}. The contour plots in Fig. 6 correspond (from left to right) to the main dynamical vacuum model RVMc and  the alternative models $Q_m$ and  $Q_{\CC}$ discussed in section 3.3. It is interesting to see that the DVM'a are in all cases favored with respect to the $\CC$CDM since the $\nu_i>0$ region is clearly projected in all the contour plots (in contrast to the $\CC$CDM case $\nu_i=0$). But even more interesting is to realize that the more recent LSS data (especially in connection to the $f(z)\sigma_8(z)$ observable for linear structure formation\,\cite{GilMarin2016}) do greatly enhance the quality fit of the DVM's versus the $\CC$CDM as compared to the older LSS data. Once more  the confidence level by which the vacuum parameters $\nu_i$ are fitted to be non null (and hence departing from the $\CC$CDM) is at the  $\sim 4\sigma$ level for models RVMc and $Q_m$. The most conspicuous one is RVMc, in which $\nu$ is nonvanishing at $4.3\sigma$ c.l. (cf. Table 2).

%
%

\subsection{Akaike and Bayessian criteria}\label{sect:AkaikeBayes}

The statistical results, in particular the relative quality of the fits from the various models, can be reassessed in terms of the time-honored Akaike and Bayessian information criteria, AIC and BIC\,\cite{Akaike,Schwarz}.


\begin{table}
\begin{center}
\caption{\scriptsize
Best-fit values for the $\CC$CDM and the three dynamical vacuum models (DVM's) with conservation of baryons and radiation, including their statistical  significance ($\chi^2$-test and Akaike and Bayesian information criteria, AIC and BIC). Once more the  $\Delta$AIC and $\Delta$BIC increments clearly favor the dynamical DE options. The RVMc and $Q_{dm}$ are particularly favored ($>4\sigma$ c.l.). Our fit is based on the same SNIa+BAO+$H(z)$+LSS+CMB data set as in Tables 1 and 2. The specific fitting parameter for each DVM is $\nueff=\nu $ (RVMc), $\nu_{dm}$($Q_{dm}$), $\nu_{\CC}$($Q_{\CC}$). The result for the XCDM is of course the same as in Tables 1 and 2 and is included here to ease the comparison with the DVM fitting results.}

\label{Table2}
\begin{scriptsize}
\resizebox{1\textwidth}{!}{
\begin{tabular}{| c | c |c | c | c | c | c | c | c | c |}
\multicolumn{1}{c}{Model} &  \multicolumn{1}{c}{$h$} &  \multicolumn{1}{c}{$\omega_b= \Omega_b h^2$} & \multicolumn{1}{c}{{\small$n_s$}}  &  \multicolumn{1}{c}{$\Omega_m$}&  \multicolumn{1}{c}{{\small$\nu_i$}}  & \multicolumn{1}{c}{$\omega$} &
\multicolumn{1}{c}{$\chi^2_{\rm min}/dof$} & \multicolumn{1}{c}{$\Delta{\rm AIC}$} & \multicolumn{1}{c}{$\Delta{\rm BIC}$}\vspace{0.5mm}
\\\hline
{\small $\CC$CDM} & $0.693\pm 0.003$ & $0.02255\pm 0.00013$ &$0.976\pm 0.003$& $0.294\pm 0.004$ & - & -1 &  90.44/84 & - & -\\
\hline
XCDM  &  $0.670\pm 0.007$& $0.02264\pm 0.00014 $&$0.977\pm0.004$& $0.312\pm 0.007$& - & $-0.916\pm0.021$ &  74.91/83 & 13.23 & 11.04 \\
\hline
RVMc  & $0.676\pm 0.005$& $0.02231\pm 0.00014$&$0.965\pm 0.004$& $0.303\pm 0.005$ & $0.00165\pm 0.00038$ & -1 &  70.32/83 & 17.82 & 15.63 \\
\hline
$Q_{dm}$  &  $0.677\pm 0.005$& $0.02229\pm 0.00015 $&$0.964\pm0.004$& $0.303\pm 0.005 $ & $0.00228\pm 0.00054 $ & -1 &  71.19/83  & 16.95 & 14.76 \\
\hline
$Q_\CC$  &  $0.692\pm 0.004$& $0.02229\pm 0.00016 $&$0.966\pm0.005$& $0.297\pm 0.004$ & $0.00671\pm 0.00246$ & -1 &  83.08/83 & 5.06 & 2.87 \\
\hline
 \end{tabular}}
 \end{scriptsize}
\end{center}
\end{table}


These information criteria are extremely useful for comparing different models in competition. The reason is that the models having more parameters have a larger capability to adjust a given set of data, but of course they should be penalized accordingly. In other words, the minimum value of  $\chi^2$ should be appropriately corrected so as to take into account this feature. This is achieved through the AIC and BIC estimators. They are defined as follow\,\cite{Akaike,Schwarz,Burnham}:
\be
{\rm AIC}=\chi^2_{\rm min}+\frac{2nN}{N-n-1}
\ee
and
\be
{\rm BIC}=\chi^2_{\rm min}+n\,\ln N\,,
\ee
where $n$ is the number of independent fitting parameters and $N$ the number of data points.
The larger are the differences $\Delta$AIC ($\Delta$BIC) with respect to the model that carries smaller value of AIC (BIC) -- the DVM's here -- the higher is the evidence against the model with larger value of  AIC (BIC) -- the $\CC$CDM.
For $\Delta$AIC and/or $\Delta$BIC in the range $6-10$ we can speak of ``strong evidence'' against the $\CC$CDM, and hence in favor of the DVM's. Above 10, we  are entitled to claim ``very strong evidence''\,\cite{Akaike,Schwarz,Burnham} in favor of the DVM's.
Specifically,  Tables 1 and 2 render $\Delta$AIC$\gtrsim15$ and $\Delta$BIC$\gtrsim15$ virtually in all cases for the type-A, type-G, RVMc and $Q_{dm}$. The results are outstanding since for all these models \emph{both} AIC and BIC peak very strongly in the same direction. Thus, these DVM's are definitely more favored than the $\CC$CDM, and the most conspicuous one is the RVMc.

We conclude that the wealth of cosmological data at our disposal currently suggests that the hypothesis $\CC=$const. despite being the simplest may well not be the most favored one. The absence of vacuum dynamics is excluded at $4\sigma$ c.l. as compared to the best DVM's considered here. The strength of this statement is riveted with the firm verdict of Akaike and Bayesian criteria. Overall we have collected a fairly strong statistical support of the conclusion that the SNIa+BAO+$H(z)$+LSS+CMB cosmological data do favor a mild dynamical vacuum evolution\,\footnote{See also the upcoming work \cite{PRDcompanion} for a subsequent reanalysis of the DVM's taking into account the last updated results from Ref.\cite{GilMarin2016}. Conclusions are essentially unchanged.}.


\begin{figure}[t]
\centering
\includegraphics[angle=0,width=0.90\linewidth]{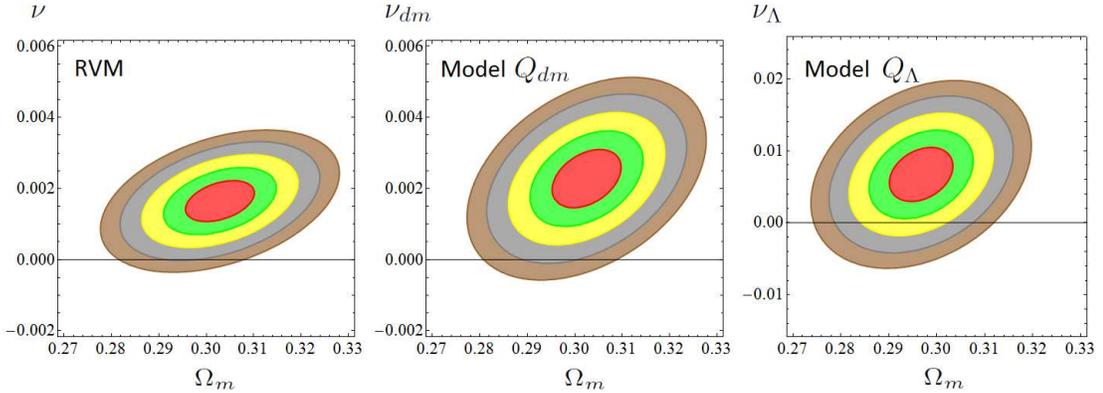}
\caption{\label{fig:CLRVMandQdm}%
\scriptsize {Likelihood contours in the $(\Omega_m,\nu_i)$ plane for the DVM's with conserved baryons and radiation, after marginalizing over the rest of the fitting parameters indicated in Table 2. The $\CC$CDM ($\nu_i=0$) appears disfavored once more at $\sim 4\sigma$ c.l. for the RVMc (denoted in the figure as RVM) and $Q_{dm}$, and at $\sim 2.5\sigma$ c.l. for the $Q_\CC$. As in Figs. 2 and 3, subsequent marginalization over $\Omega_m$ increases slightly the c.l. and renders fitting values with a statistical
significance of $\gtrsim4.2\sigma$ for the RVMc and $Q_{dm}$.}
}
\end{figure}
%


\section{RVM's and the cosmic time evolution of masses and couplings}\label{sec:RunMass}

The framework outlined in the previous sections suggest that owing to a small interaction with vacuum the matter density of the Universe might not be conserved during the cosmic expansion and in such case it would obey an anomalous conservation law. However, if there is a feedback of matter with the cosmic vacuum it would open a new window into the domain of the time variation of the fundamental constants of Nature\footnote{For a summarized introduction, see e.g.\,\cite{Preface}. The idea traces back to early proposals in the thirties on the possibility of a time evolving gravitational constant $G$ by Milne\,\cite{Milne1935} and the suggestion by Dirac of the large number hypothesis \cite{Dirac1937}, including the ideas by Jordan that the fine structure constant $\alpha_{\rm em}$ together with $G$ could be both space and time dependent\,\cite{Jordan1937,JordanBook}.}. Such window would permit a limited, but nonvanishing, exchange of information between the two widely different worlds of Quantum Physics and General Relativity. In point of fact it could be considered as a kind of living fossil  (i.e. what is left at present) of the glorious unification of QP and GR in the remote past, and therefore the late testimony of such epoch in the very early Universe.  We have elsewhere called that subtle interplay:``the micro and macro connection''\,\cite{FritzschSola2015,FritzschSola2012}. If by some chance the physical laws still use a narrow passage of this sort to communicate the two ostensibly divorced worlds of QP and GR, there might be still hope for a real understanding of the grand picture of the Cosmos!

A large number of studies have been undertaken in our days from different perspectives on the possible variation of the fundamental ``constants'', sometimes pointing to positive observational evidence\,\cite{alphat1} but often disputed by alternative observations\,\cite{alphat2} --  see e.g. the reviews \cite{Barrow2010,Fritzsch2009,ConstantsNature1,ConstantsNature2,ConstantsNature3}.
Of essential importance is to count on a consitent theoretical framework. The DVM's, and in particular the subclass of RVM's might offer new clues for a possible explanation of the potential reality of these subtle and often evanescent effects, which are being constantly scrutinized by means of highly sophisticated lab experiments and accurate astrophysical observations\,\cite{ConstantsNature4,SpecialIssueMPLA}.

In the simplest RVM case (\ref{eq:RVMvacuumdadensity}) the anomalous matter conservation law takes on the form given in  Eq.\,(\ref{MatterEvolution}), which is concomitant with a mild vacuum evolution of the sort (\ref{VacuumEvolution}). However, here we wish to reinterpret the anomalous matter conservation law in a different way\,\cite{FritzschSola2012,FritzschSola2015}, see also \cite{FritzschSola2014,Sola2014}. Rather than assuming that the presence of a nonvanishing $\nu$ is related to an anomalous conservation of the number density of particles, we may conjecture that it is associated to the nonconservation of the particle masses themselves. In other words, we suppose that while there is a normal dilution of the particle number density with the expansion for all the particle species, i.e. $n_i=n_i^0\,a^{-3}=n_i^0(1+z)^3$, the corresponding mass values $m_i$ are not preserved throughout the cosmic expansion:
\begin{equation}\label{eq:mz}
m_i(z)=m_i^0\,(1+z)^{-3\nu_i}\,,
\end{equation}
$m_i^0$ being the current values ($z=0$) of the particle masses, and the various $\nu_i$ are the different anomaly indices for the non-conservation of each species. This was the point of view adopted in\,\cite{FritzschSola2012}. Notice that $\rho_{m_i}$ in Eq.\,(\ref {MatterEvolution}), i.e. the mass density for the ith-species of particles, is indeed equal to $n_i(z) m_i(z)$, with  $ m_i(z)$ given by (\ref{eq:mz}) and $\rho_{m_i}^0=n_i^0\,m_i^0$.  In the remaining of this section it will be necessary to distinguish among the different values of $\nu_i$ for each particle species, but for simplicity we shall differentiate only between the two large families of baryons and DM particles to which we will attribute the generic indices $\nu_b$ and $\nu_X$ respectively\,\footnote{The anomaly mass indices $\nu_i=\nu_b,\nu_X$ here should not be confused with the vacuum parameters $\nu_i$ introduced for the different DVM's in the previous sections. However, once a DVM is selected, e.g. the canonical RVM defined by (\ref{eq:RVMvacuumdadensity}), we have a collection of anomaly indices $\nu_i$ for the different particle species which are related to the RVM vacuum parameter $\nu$ through (\ref{mass3}) below. } ($X$ denoting here the generic particles contributing to the DM).  Formulated in this fashion, such scenario is in principle closer to the general RVM one discussed in section 3.1 rather than to the more specific RVMc framework discussed in section 3.2, in which baryons and radiation were assumed to be conserved. However, as we will discuss below, the observational measurements actually suggest that $\nu_b\ll\nu_X$ and, therefore, in practice we will effectively stay quite close to the RVMc, namely to that very successful dynamical vacuum model capable of surpassing the  $\CC$CDM ability to describe the cosmological observations at $\sim 4\sigma$ c.l. (cf. sect. 4). This may be viewed as an additional motivation for such a particular realization of the RVM since it is naturally compatible both with the cosmological observations as well as with the current bounds on the possible variation of the fundamental constants.

The above interpretation leads to a peculiar ``Weltanschauung'' of the physical world, in which basic quantities of the standard model (SM) of strong and electroweak interactions, such as the quark masses, the proton mass and the quantum chromodynamics (QCD)
scale parameter, $\Lambda_{QCD}$, might not be conserved in the course of the cosmological evolution, see \cite{FritzschSola2012,FritzschSola2015}.
Let us take, for example, the proton mass, which is given as follows:
$m_p = c_{QCD} \Lambda_{QCD} + c_u m_u + c_d m_d + c_s m_s + c_{em} \Lambda_{QCD}$, where $m_{u,d.s}$ are the quark masses and the last term represents the electromagnetic (em) contribution.  Obviously the leading term is the first one, which is due to the strong binding energy of QCD. Thus, the nucleon mass can be expressed to within very good approximation as $m_p \simeq c_{QCD} \LQCD \simeq 938 MeV$, in which
$c_{QCD}$ is a non-perturbative coefficient. The masses of the light quarks $m_u$,
$m_d$ and $m_s$ also contribute to the the proton mass, although with less than 10$\%$ and can therefore be neglected for this purpose. It follows that cosmic time variations of the proton mass are essentially equivalent to cosmic time variations of the QCD scale parameter:
\begin{equation}\label{eq:mpQCD}
\frac{\dot{m}_p}{m_p}\simeq \frac{\dot{\Lambda}_{\rm QCD}}{\LQCD}\,.
\end{equation}
On the other hand, the QCD scale parameter is related to the strong coupling constant $\alpha_s = g^2_s/4 \pi$ as follows (at 1-loop order):
\begin{equation}\label{eq:alphasLQCD}
 \alpha_s(\mu_R) = \frac{4 \pi}{(11-2 n_f/3) \ln( \mu^2_R/\Lambda^2_{QCD})}\,.
\end{equation}
Here $\mu_R$ is the renormalization scale, $n_f$ the number of quark flavors and $\Lambda_{QCD}$  = 217 $\pm$ 25 MeV the measured value of the QCD scale paramete. Now, if there is a feedback between the micro and macro world, we should be ready to admit the possibility that when we consider QCD in the context of a FLRW expanding universe the value of the proton mass, and hence of $\Lambda_{QCD}$, need not remain constant anymore. The possible variation of $\Lambda_{QCD}$ should, of course, be relatively small. In the kind of scenario described in the previous sections the running of the cosmological parameters is settled by the cosmic scale $\mu_c\equiv H$. Therefore, if there is a feedback between the micro and macro world, the scale $H$ should also define the natural rhythm of variation of the subnuclear parameters, up to dimensionless coefficients which can vary from one parameter to the next. On these grounds we expect that the cosmic variation of the QCD scale is regulated by the Hubble rate, i.e. $\Lambda_{QCD} = \Lambda_{QCD}(H)$. As a result the strong coupling constant $\alpha_s$ becomes a function not only of the conventional renormalization scale  $\mu_R$  but also of the cosmic scale $\mu_c\equiv H$ . Since $H=H(z)$ is a function of the cosmological redshift, we can write  $\alpha_s=\alpha_s(\mu_R, z)$ and
from Eq.\,(\ref{eq:alphasLQCD}) we find that the relative variation of the strong coupling constant with the Hubble rate is related with the corresponding variation of $\Lambda_{QCD}$ in the following manner (at one-loop):
\begin{equation}
\label{alpha_s}
\frac{1}{\alpha_s} \frac{d \alpha_s(\mu_R,z)}{dz} = \frac{1}{\ln(\mu_R/\LQCD(z))} \Big[ \frac{1}{\LQCD(z)}\frac{d \LQCD(z)}{dz} \Big]\,.
\end{equation}
If the QCD coupling constant $\alpha_s$ or the QCD scale parameter $\Lambda_{QCD}$ undergo some cosmological shift, the nucleon masses as well as the masses of the atomic nuclei would also red shift along with $\Lambda_{QCD}$.

\begin{figure}[t]
\begin{center}
\label{Hznu1}
\includegraphics[width=3.8in, height=2.7in]{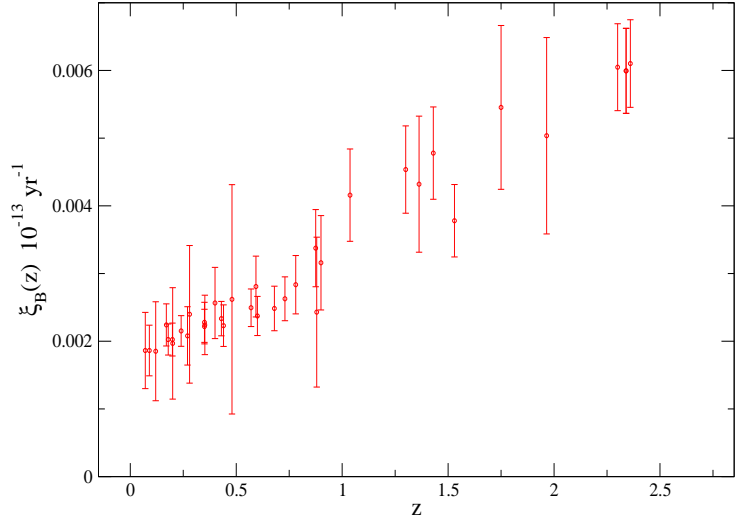}
\caption{Contribution from baryons
to the total mass drift rate. We plot the function $\xi_B(z)$ from (\ref{indices}) for $\nu_b=10^{-5}$ using the $H(z)$ data points at different redshifts (see text).}
\end{center}
\end{figure}


\begin{figure}[t]
\begin{center}
\label{Hznu2}
\includegraphics[width=3.8in, height=2.7in]{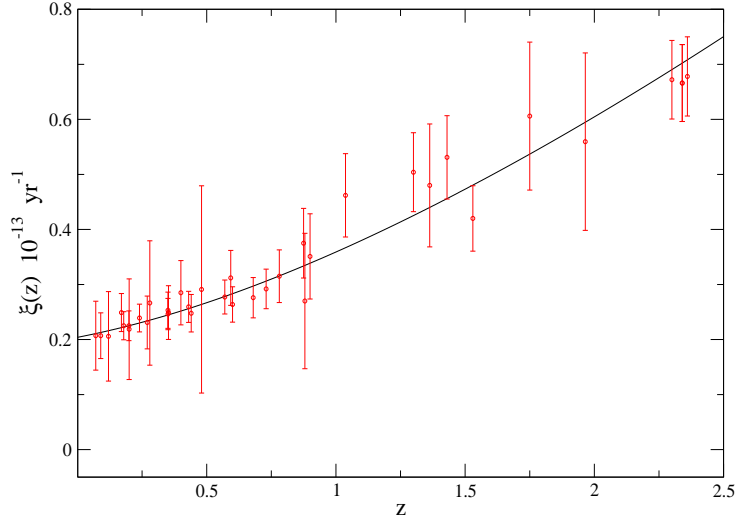}
\caption{The total mass drift rate $\xi(z)$, given by Eq.\,(\ref{xifunction}) as predicted by the RVM,
as a function of the redshift and within the same conditions as in the previous figure. Here $\nu_X=10^{-3}$. This value essentially saturates the fitted $\nu$ from cosmology in sect. 4 (recall that $\nu_b\ll\nu$, see the text).}
\end{center}
\end{figure}



\subsection{Cosmic drift of particle masses}

In this study we attribute the cosmic variation of the particle masses to the energy exchange with the cosmic vacuum according to the RVM framework outlined in the previous section, in which a possible additional variation of the gravitational constant may also concur.
In order to estimate quantitatively these effects within the RVM, we take as a basis the numerical fit estimates obtained in section 4 using the known data on SNIa+BAO+H(z)+LSS+BBN+CMB.  Among these observational sources we use 36 data points on the Hubble rate $H(z)$ at different reshifts in the range $0 < z \leq 2.36$, as compiled in \cite{Hdata,RatraFarooq2013}. These data will play a significant role in our aim to estimate the possible temporal evolution of the particle masses, both for baryons and dark matter, as we shall show next.

In order to estimate the variation of the particle masses within the RVM, we start by noting that the matter density of the universe can be approximated as $\rho_m \simeq n_p m_p + n_n m_n + n_{X} m_{X}$,
where we neglect the leptonic contribution and the relativistic component (photons and neutrinos).
Here $n_p , n_n , n_{X}$ ($m_p , m_n , m_{X}$) are the number densities (and corresponding masses) of protons, neutrons and dark matter (DM) particles $X$, respectively. Assuming that the anomalous mass density law in eq. (\ref{MatterEvolution}) is a direct reflect of the change of the particle masses, the anomalous fractional mass density time variation of the Universe can be estimated as follows\,\cite{FritzschSola2012,FritzschSolaNunes2016}:
\begin{equation}\label{eq:reltimeMdensityUniv1}
\frac{\delta\dot{\rho}_m}{\rho_m}\simeq
\frac{n_p\,\dot{m}_p+n_n\,\dot{m}_n+n_X\,\dot{m}_X}{n_X\,m_X}\,\left(1-\frac{\Omega_b}{\Omega_{dm}}\right)\,,
\end{equation}
where $\Omega_b$ and $\Omega_{dm}$ represent as usual the fractional density of baryons and DM particles with respect to the critical density, respectively. Of course the total $\Omega_m$ is equal to the sum $\Omega_b+\Omega_{dm}$. In the above formula we have used the fact that  $\Omega_b/\Omega_{dm}=\rho_b/\rho_{dm}$ with $\rho_b=n_pm_p+n_nm_n$ and $\rho_{dm}=n_Xm_X\equiv\rho_X$ is the density of DM particles (which we have called generically $X$ for simplicity, hence the notation $\rho_X$).  Equation (\ref{eq:reltimeMdensityUniv1}) can be further expanded as follows. Let us take  $m_n = m_p \equiv m_p$ so that $\rho_b=(n_p+n_n)\, m_p$, and assume $\dot{m}_n = \dot{m}_p\equiv\dot{m}_p$. Since $n_n/n_p$ is of order $10\%$ after the primordial nucleosynthesis and at the same time $\Omega_b/\Omega_{dm}={\cal O}(10^{-1})$,  we may neglect the product of these two terms or any higher power of them. In this way we arrive at the following simple expression:\cite{FritzschSola2012}
\begin{equation}
\label{mass1}
 \frac{\delta \dot\rho_m}{\rho_m} \simeq \Big(1- \frac{\Omega_b}{\Omega_{dm}} \Big) \Big(\frac{\Omega_b}{\Omega_{dm}} \frac{\dot{m}_p}{m_p} + \frac{\dot{m}_{X}}{m_{X}} \Big)\,.
\end{equation}
The vacuum parameter of the RVM  can now be related with the individual mass variations of baryons and dark matter. Using Eq. (\ref{MatterEvolution}), the anomalous fractional mass density time variation  can  be written  $\delta \dot\rho_m/ \rho_m \simeq 3 \nu H$, within the linear approximation of the small parameter $\nu$ and for moderate values of the redshift. Replacing this expression on the \textit{l.h.s} of Eq.\,(\ref{mass1}) we can rephrase it in the following  way:
\begin{equation}
\label{mass2}
\frac{\nu }{1 - \Omega_b/\Omega_{dm}}=\frac{\Omega_b}{\Omega_{dm}} \nu_b + \nu_{X}\,,
\end{equation}
where we have introduced the equations that define the corresponding mass drift rates for baryons and DM particles:
\begin{equation}
\label{indices}
\xi_B(t)\equiv\frac{\dot{m_p}}{m_p} = 3 \nu_b H, \,\,\,\,\,\,\,\,\,\,\,\,\,\,\,\,\,  \xi_X(t)\equiv\frac{\dot{m}_{X}}{m_{X}}= 3 \nu_{X}H\,,
\end{equation}
 in which $\nu_b$ and $\nu_{X}$ are the anomaly indices  for the evolution of the baryons and DM, in accordance to our discussion in the beginning of this section.
Assuming that the anomaly indices for matter non-conservation are constant we can easily integrate these equations in terms of the redshift variable.  Using the by now familiar relation $\dot{m}=aH dm/da=-(1+z)\,H\,dm/dz$, we find:
\begin{equation}\label{eq:nubnuX}
m_p(z)=m_p^0\,(1+z)^{-3\nu_b}\ \ \ \ \ \ \ \ \ \ m_X(z)=m_X^0\,(1+z)^{-3\nu_X}\,,
\end{equation}
respectively for baryons (essentially the proton as the only stable baryon) and DM particles.  As we can see, these equations in fact lead to the generic form (\ref{eq:mz}).

The total drift rate from the time variation of the masses of all heavy and stable particles in the Universe (baryons + dark matter) reads
\begin{equation}
\label{xifunction}
\xi(t) = \xi_B(t)+\xi_X(t) = 3H (\nu_b + \nu_{X}).
\end{equation}
The drift rates are of course functions of time and redshift. Usually the relation with the variation of a particular mass $m_i$, baryon or DM, within a cosmological span of time $\Delta t\sim H^{-1}$ is approximated in a linear way, i.e. one assumes that on average the time variation of the mass was the same during the time interval $\Delta t$. In this way we have
\begin{equation}\label{eq:variationmi}
\frac{\dot{m}_i}{m_i}\simeq \frac{\Delta m_i}{m_i\,\Delta t}\simeq \frac{\Delta{m_i}}{m_i}\,H \ \ \ \rightarrow\ \ \  \frac{\Delta{m_i}}{m_i}\simeq 3\nu_i\,.
\end{equation}
Thus, the anomaly mass indices $\nu_i$ account for the typical mass variation of a given particle species (baryons or DM particles) in a cosmological span of time.

Since, as mentioned, $\Omega_b/\Omega_{dm}\simeq 0.1$, we can neglect the square of this quantity  and rewrite (\ref{mass2}) in the more compact form
\begin{equation}
\label{mass3}
\nu =\frac{\Omega_b}{\Omega_{dm}} \left(\nu_b-\nu_{X}\right) + \nu_{X}\,.
\end{equation}
Interestingly, this equation can be checked experimentally, for $\nu$ can be fitted from the cosmological observations based on SNIa+BAO+H(z)+LSS+BBN+CMB data, as indicated in the previous section, where it was found to be of order $10^{-4}-10^{-3}$, whereas $\nu_b$ can be determined from specific astrophysical and lab experiments trying to measure the possible time evolution of the ratio $\mu=m_p/m_e$\,\cite{ConstantsNature1}. Thus, if the equation  (\ref{mass3}) must be satisfied, we can verify if the DM part $\nu_{X}$ (which, of course, cannot be measured in an isolated way) plays or not a significant role in it.

From a rich variety of experimental situations both from astrophysical observations and direct lab measurements\,\cite{ConstantsNature1,ConstantsNature2,ConstantsNature3} (most of them compatible with a null test) one finds that $\Delta\mu/\mu$ is at most in the ballpark of ${\cal O}(1-10)$ parts per million (ppm). Let us note that
\begin{equation}\label{eq:Deltamuovermu}
\frac{\Delta\mu}{\mu}=\frac{\Delta m_p}{m_p}-\frac{\Delta m_e}{m_e}=3(\nu_b-\nu_{\ell})\,,
\end{equation}
where we have used Eq.\,(\ref{eq:variationmi}). The index $\nu_b$ was applied to the proton as the only stable baryon, whereas $\nu_{\ell}$ corresponds to the electron as the only stable lepton. It is usually assumed that $\nu_b\gg\nu_{\ell}$ and then $\Delta\mu/\mu\simeq\Delta m_p/m_p$. In this case the aforementioned limit on $\Delta\mu/{\mu}$ would imply $\nu_b\sim 10^{-5}$ at most. However, a more symmetric option (which cannot be ruled out at present) is that the two indices $\nu_b$ and $\nu_{\ell}$ can be close to each other. In such case both could be of order $10^{-4}$ and numerically similar. Note that this option would still be compatible with the approximate bounds on ${\Delta\mu}/{\mu}$ of at most 10 ppm. It is fair to keep in mind these two possibilities in our analysis.  Both of them, however, lead to $\nu_X\sim\nu\sim 10^{-3}$ via Eq.\,(\ref{mass3}), what clearly points to the crucial role of the DM contribution to explain the bulk of the mass drift rate in the Universe (cf. Figs. 7 and 8).

In general we have the anomaly law (\ref{eq:mz}) for all species of particles, whether baryons, leptons or DM particles ( $i = b, \ell, X$), and from it we may compute the relative variation of the mass with the redshift:
\begin{equation}
\label{mass}
 m_i(z)=m_{i0}(1+z)^{-3\nu_i}  \,\,\,\,  \longrightarrow   \,\,\,\,  \frac{\Delta m_i(z)}{m_i} \simeq -3 \nu_i \ln(1+z)\,.
\end{equation}
Here we have defined $\Delta m_i(z) = m_i(z) - m_{i0}$, with $m_{i0}\equiv m_i(z=0)$ the current value, so that the $\Delta m_i(z)/m_i$  above gives the fractional variation of the mass at redshift $z$ as compared to the current mass, assuming that $z$ is not very large such that $|\nu_i|\ln(1+z)<1$ holds good. In the framework outlined in the previous sections these particle mass changes with time are possible thanks to the interaction with the dynamical vacuum and/or the evolution of the gravitational constant. It is well-known that Brans-Dicke (BD) type models of gravity can also provide the time evolution of the gravitational coupling\,\cite{BD}. Interestingly enough, it has recently been shown\,\cite{Elahe2} that generalized BD-type models can also provide the time variation of the particle masses in a way which is fully consistent with the variation of $G$. These same models actually provide a possible fundamental origin of the Higgs potential in the BD-gravity framework\,\cite{Elahe2}.

\subsection{Cosmic drift of the fine-structure constant}\label{sect:alphaem}

Interestingly, there are indications of a possible variation (decrease) of the fine-structure constant at high redshift, as well as a possible spatial variation, (see \cite{alpha0} and references therein, as well as the reviews\,\cite{ConstantsNature1,ConstantsNature2,ConstantsNature3}). We will address here this topic from the point of view of the implications of the running vacuum energy density throughout the cosmic history, i.e. the RVM picture. Let us note that in the electroweak sector of the SM one cannot establish a direct connection between the time evolution of the electroweak couplings $g$ and $g'$ from $SU(2)\times U(1)$ to that of the particle masses masses since there is no analogue in this sector of the QCD scale parameter $\LQCD$. Notwithstanding, it is still possible to relate them to $\LQCD$ in an indirect way if we use the hypothesis of Grand Unification of the SM couplings at a very high energy scale.  We will focus here on the fine-structure constant $\alpha_{\rm em}=e^2/4\pi$ and its correlated time-evolution with the strong coupling counterpart $\alpha_{s}=g_s^2/4\pi$, and ultimately with the time evolution of $\LQCD$ and $\mu=m_p/m_e$.
\\

As it is well-known, ia Grand Unified Theory (GUT), the gauge couplings, and in particular the strong coupling given in eq. (\ref{eq:alphasLQCD}), converge at the unification point with the electroweak couplings. This property can trigger a theoretical argument so as to connect the possible time variation of the electroweak  running coupling constants and the proton mass\,\cite{FritzschSola2012}.
Let $d\alpha_i/dz$ be the variation of $\alpha_i$ with the cosmological redshift $z$. Such variation is possible if we have a consistent theoretical framework supporting this possibility, such as the RVM picture described in sect.3. Each of the couplings $\alpha_i=g^2_i/4\pi$ ($g_i=g,g',g_s$)
is a function of the running scale $\mu_R$, and they follow the standard (1-loop) running laws
\begin{equation}\label{eq:GUTrunning}
\frac{1}{\alpha_i(\mu_R,z)}=\frac{1}{\alpha_i(\mu'_R,z)}+\frac{b_i}{2\pi}\ln\frac{\mu'_R}{\mu_R}\,,
\end{equation}
in which we have included the redshift variable to parametrize the cosmic evolution. If we take into account that the $\beta$-function coefficients $b_i$ of the running are constant in time and redshift, it follows that the expression $\alpha_i'(z)/\alpha_i^2(z)\equiv(d\alpha_i /dz)/\alpha^2_i$ is independent of $\mu_R$, i.e. it is a RG-invariant.
With the help of this property and our ansatz concerning the cosmological evolution of the particle masses in the RVM, it is possible to show that the running of the electromagnetic coupling $\alpha_{em}=e^2/4\pi$ is related to that of the strong coupling $\alpha_{s}$  as follows\,\,\cite{FritzschSola2012,CalmetFritzsch}:
\begin{equation}
\frac{1}{\alpha_{em}} \frac{d \alpha_{em} (\mu_R,z)}{dz} = \frac{8}{3} \frac{\alpha_{em}(\mu_R,z)}{\alpha_{s}(\mu_R,z)}\frac{1}{\alpha_{s}} \frac{d \alpha_{s} (\mu_R,z)}{dz}\,.
\end{equation}
Recalling now  Eq.\,(\ref{alpha_s}) we can reexpress the cosmic running of ${\alpha_{em}}$ in terms of the cosmic running of the QCD scale:
\begin{equation}
\frac{1}{\alpha_{em}} \frac{d \alpha_{em} (\mu_R,z)}{dz} = \frac{8}{3} \frac{\alpha_{em}(\mu_R,z)/\alpha_{s}(\mu_R,z)}{\ln(\mu_R/\Lambda_{QCD})}\frac{1}{\LQCD}\frac{d \LQCD(z)}{dz}.
\end{equation}
Evaluating the above expression at the  $Z$-boson mass scale $\mu_R = M_Z$, where both $\alpha_{em}$ and $\alpha_s$ are known with precision, it yields
\begin{equation}\label{eq:alphaemLQCD}
\frac{1}{\alpha_{em}} \frac{d \alpha_{em} (\mu_R,z)}{dz} \simeq 0.03 \frac{1}{\LQCD}\frac{d \LQCD(z)}{dz}.
\end{equation}
This equation can now be naturally linked with our discussion of the cosmic running of the particle masses considered in sect. 3. Indeed, we have seen that the proton mass receives the bulk of its contribution from $\LQCD$ through $m_p\simeq c_{QCD}\,\LQCD$ (with a negligible contribution from the quark masses and the electromagnetism). Using this expression in Eq.\,(\ref{eq:alphaemLQCD}), integrating and finally inserting the redshift dependence of the proton mass through (\ref{eq:nubnuX}), we find:
\begin{equation}\label{eq>alphaemz}
\alpha_{\rm em}(z)\simeq \alpha_{\rm em}^0\left(\frac{m_p(z)}{m_p^0}\right)^{0.03}=\alpha_{\rm em}^0\,(1+z)^{-0.09\nu_b}\,,
\end{equation}
where $\alpha_{\rm em}(z)$ represents the value of the fine structure constant at redshift $z$ at a fixed value of $\mu_R$, and $\alpha_{\rm em}^0(z)$ is its current value ($z=0$). Being $\nu_b$ a small parameter, which is related to the fitted value $\nu\sim 10^{-3}$ through (\ref{mass3}), we can estimate the fractional variation of  $\alpha_{em}$ with the redshift as follows:
\begin{equation}\label{eq:alphavariation}
\frac{\Delta\alpha_{\rm em}(z)}{\alpha_{\rm em}}\simeq -0.09\,\nu_b\,\ln(1+z)\,.
\end{equation}
By analogy  with Eq.\,(\ref{mass}), we learn that the effective running index of the electromagnetic coupling, $\nu_{\rm em}$, is some $30$ times smaller than that of the baryonic index and with opposite sign, in other words: $\nu_{\rm em}\simeq -0.03\,\nu_b$.

\begin{figure}[t]
\begin{center}
\label{fig:alphavalues}
\includegraphics[width=3.8in, height=2.7in]{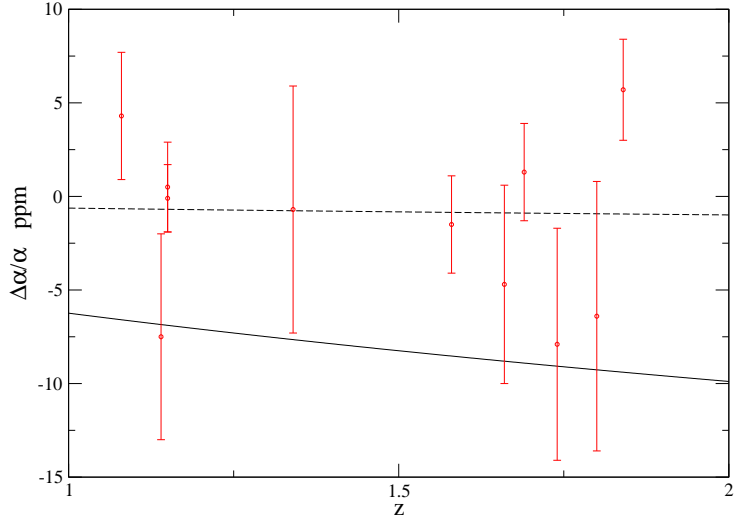}
\caption{Data points on the relative variation $\Delta\alem/\alem$ at different redshifts (in ppm). For the exact numerical values and observational references, see Table 1 of \cite{FritzschSolaNunes2016}.
The solid and dashed lines correspond to the theoretical combined RVM-GUT prediction based on formula
(\ref{eq:alphavariation}) for the values $\nu_b=10^{-4}$ and $10^{-5}$, respectively.
The tendency of the data, roughly suggesting smaller values of $\alem$ at higher redhifts, is correctly described by the theoretical curves,
although the current observational errors are still too large to extract firm conclusions.}
\end{center}
\end{figure}

At this point it is worthwhile to note that the observational results\,\cite{ConstantsNature1,ConstantsNature2,ConstantsNature3}, whether from astrophysical or lab measurements, can be accounted for (in order of magnitude) within the RVM (essentially within the RVMc since $\nu_b\ll\nu$)  in combination with the GUT hypothesis. Indeed, the  theoretical RVM prediction falls correctly within the order of magnitude of the typical measurements in Fig.\,9, provided $\nu_b$ lies in the range from $10^{-4}$ to $10^{-5}$.  This follows from Eq.(\ref{eq:alphavariation}), which, roughly speaking, says that the RVM prediction is of order $\Delta\alpha_{\rm em}/\alpha_{\rm em}\sim -0.09\nu_b$ up to log corrections in the redshift. In Fig.\,9 we have superimposed the exact theoretical prediction $\alpha_{\rm em}(z)$  according to the formula (\ref{eq:alphavariation}). We can see that, notwithstanding the sizeable error bars, the trend of the measurements suggests a decrease of $\alpha_{\rm em}$ with the redshift (as there are more points with $\Delta\alem<0$ than points with $\Delta\alem>0$). This behavior has been previously noted in the literature and is roughly in accordance with our theoretical curves  in Fig.\,9. But of course we need more precise measurements to confirm the real tendency of the data, as the errors are still too large and no solid conclusion is currently possible.

It is remarkable that the correct order of magnitude for the maximal possible value of $\nu_b$, which we have obtained from the direct $\Delta\alem/\alem$ observations (viz. $\nu_b\sim 10^{-4}-10^{-5}$), does coincide with the result inferred from our previous considerations on the alternative observable $\Delta\mu/\mu$, Eq.\,(\ref{eq:Deltamuovermu}). This is because the RVM in combination with the GUT framework neatly predicts the relation $\nu_{\rm em}\simeq -0.03\,\nu_b$ (up to a logarithmic correction with the redshift).

The general conclusion that ensues from our analysis is that the baryonic index $\nu_b$ in Eq.\,(\ref{mass3}) is definitely subdominant as compared to the dark matter one, i.e. $|\nu_b|\ll|\nu_X|$. This is consistent with our assumption of approximate baryon mass conservation in the RVMc, the model studied in section 3.2. Being $|\nu_b|$ very small it is clear that $\nu_X$ must be of order of the total matter index $\nu\sim 10^{-3}$ fitted from the overall cosmological observations. In other words, we find once more that it must be the DM component that provides the bulk of the contribution to the time variation of masses in the Universe. This fact was not obvious a priori, and should not be confused with the overwhelming abundance of DM as compared to baryons. After all the large amounts of DM could simply remain passive and not evolve at all throughout the cosmic expansion.  For example, if the best fit value of  $\nu$ would have been found in the ballpark of  $10^{-5}$, equation (\ref{mass3}) could have still been fulfilled with $\nu_X<<\nu_b\sim 10^{-4}$ and yet this would be roughly compatible with the measurements of $\Delta\alpha_{\rm em}/\alpha_{\rm em}$.  However, the fact that the value of $\nu$ (obtained from the overall cosmological fit to the data within the RVM, cf. sect. 4) comes out significantly larger than the baryonic index $\nu_b$ has nontrivial consequences and provides an independent hint of the need for  (time-evolving) dark matter. The fact that this same conclusion can be inferred both from the analysis of $\Delta\mu/\mu$ and $\Delta\alem/\alem$, which become correlated in this theoretical framework, does reinforce the RVM class of dynamical vacuum models and places the contribution from the dark matter component to the forefront of our considerations concerning the total mass drift rate in the Universe\,\cite{FritzschSola2012,FritzschSola2015,FritzschSolaNunes2016}.

\section{Conclusions}

The idea that the cosmic vacuum should be dynamical in an expanding Universe is not only a theoretically appealing possibility but also a phenomenologically preferred option. Here I have reviewed the current level of evidence supporting this statement. To this end, several dynamical vacuum models (DVM's) have been discussed and confronted with a large set of data involving the cosmological observables  SNIa+BAO+$H(z)$+LSS+BBN+CMB, and we have found that the DVM's can perform significantly better than the $\CC$CDM. In other words, the dynamical vacuum models considered here do challenge in a rather bold manner the $\CC$CDM in its ability to describe the main set of cosmologically significant observations. Among the DVM's the subclass of running vacuum models (RVM's) stands out.
The excellent current status of the various DVM's (most particularly the RVMc, in which baryon number and radiation are conserved) is to be emphasized. It can be easily appreciated from the summary plots displayed in Figs. 2,3 and 6, where the $\CC$CDM model is comparatively disfavored. The data are currently able to discriminate between the nul value of the vacuum parameter $\nu=0$ (corresponding to a rigid $\CC=$ const. as in the $\CC$CDM) and positive nonvanishing values of $\nu\sim 10^{-3}$ at a confidence level of $4\sigma$ c.l. ($>99.9\%)$ for the main DVM's. In fact the analysis of WMAP9, Planck 2013 and the recent Planck 2015 data, all of them point to the same conclusion with different levels of evidence in proportion to the accuracy of each one of these experiments. We find that the  RVM class of dynamical models (\ref{eq:RVMvacuumdadensity}), including the generalized ones (\ref{eq:A2}), are strongly preferred as compared to the concordance $\CC$CDM. The precise meaning of ``strong evidence'' can be carefully quantified in terms of Akaike and Bayesian statistical criteria for model comparison. These criteria render more than 15 points of difference in favor of the RVM's (in a scale where already 10 points means ``very strong'' evidence) against a rigid $\CC$. It seems that the phenomenological support to the RVM's, in detriment of the $\CC$CDM, is fairly robust in the light of the current data.

Finally, the RVM framework has an additional bonus. It can also provide an explanation for the possible (slow) time-evolution of the fundamental constants of Nature. This is a field which probably holds many surprises in the future\,\cite{Preface,SpecialIssueMPLA}. The natural impact from the RVM on this issue occurs thanks to the cosmological exchange of energy between vacuum, matter and the possible interplay with the Newtonian coupling $G$. Because $\mu_c\sim H$ is the natural scale in the cosmological renormalization group running\,\cite{JSPReview2013}, the RVM predicts that the associated rhythm of change of the fundamental constants (such as couplings, masses and vacuum energy density) should naturally be as moderate as dictated by the value of the Hubble rate at any given epoch. The RVM thus sets the time scale $1/H$ and a characteristic rhythm of variation of order $\dot{\cal P}/{\cal P}\sim H_0\,\Delta{\cal P}/{\cal P}\lesssim \left(\Delta{\cal P}/{\cal P}\right)\,10^{-10}$yr$^{-1}$ for any parameter ${\cal P}$, hence in the right ballpark of the usual bounds on the possible running of the fundamental constants. Typically $\Delta{\cal P}/{\cal P}\lesssim 10^{-4}-10^{-3}$ over a cosmological span of time, which depends on the monitored parameter ${\cal P}=\CC, G, m_i,\alpha_{\rm em}, \alpha_s, \Lambda_{\rm QCD}$...\,\cite{FritzschSola2012}. A drift rate of that order emerges from the overall fit to the cosmological data (cf. section 4) and would be compatible with a drift rate of the dark matter particles of the same order of magnitude. Since the experimental bounds on the time variation of e.g. the proton-to-electron mass, $\mu=m_p/m_e$, are usually smaller by at least one order of magnitude, we conclude that it must be the time drift of the dark matter particles which compensates for the difference (cf. Figs. 7-8). This could lead to an indirect confirmation of the existence of dark matter, namely through its necessary (and faster) cosmic time evolution as compared to baryonic matter. Moreover, the fact that the baryonic matter is essentially conserved (i.e. not time evolving) is in accordance with the fact that the most successful model within the RVM clas is the RVMc, i.e. the RVM variant in which baryons and radiation are conserved (in contrast to dark matter, which exchanges energy with vacuum). We also find that this scenario is consistent with the limits on the possible variation of the fine structure constant (cf. Fig. 9), although more accuracy in these measurements is needed.

The above class of RVM scenario points, therefore, to the possibility that there is a subtle crosstalk between the atomic world and the Universe in the large, which may be on the verge of being detected. We have called it ``the micro and macro connection''\,\cite{FritzschSola2015}. It amounts to an almost imperceptible feedback between those two worlds (micro and macro) and is responsible for a mild time drifting of the ``fundamental constants'' of Nature, in a way which is perfectly consistent with the general covariance of Einstein's equations\,\cite{FritzschSola2012}. Testing these ideas might hint at the missing link between the physics of the very small and the physics of the very large\,\cite{Preface}, i.e. the (long sought-for) overarching interconnection of the subatomic quantum mechanical world with the large scale structure of the Universe. Ultimately, it might inject some more optimism for an eventual solution of the hard cosmological constant problem and the associated cosmic coincidence problem.

\vspace{0.75cm}
\textbf{Acknowledgments}
\vspace{0.30cm}

It is my pleasure to thank Prof. Harald Fritzsch for the kind invitation to this stimulating conference on LHC physics and for his collaboration on part of the work presented here. I would also like to thank Prof. K. K. Phua for inviting me
to present this contribution in the review section of IJMPA. The numerical results on the recent comparison between the RVM's and the $\CC$CDM have been obtained in different works in collaboration with  my students  Javier de Cruz P\'erez, Adri\`a G\'omez-Valent and Elahe Karimkhani. I am thankful to them, as well as to Rafael Nunes, for his collaboration on part of this analysis.
I have been supported
by FPA2013-46570 (MICINN), CSD2007-00042 (CPAN),
2014-SGR-104 (Generalitat de Catalunya),  MDM-2014-0369 (ICCUB) and by the Institute for Advanced Study of the Nanyang Technological University in Singapore.


\newcommand{\CQG}[3]{{ Class. Quant. Grav. } {\bf #1} (#2) {#3}}
\newcommand{\JCAP}[3]{{ JCAP} {\bf#1} (#2)  {#3}}
\newcommand{\APJ}[3]{{ Astrophys. J. } {\bf #1} (#2)  {#3}}
\newcommand{\AMJ}[3]{{ Astronom. J. } {\bf #1} (#2)  {#3}}
\newcommand{\APP}[3]{{ Astropart. Phys. } {\bf #1} (#2)  {#3}}
\newcommand{\AAP}[3]{{ Astron. Astrophys. } {\bf #1} (#2)  {#3}}
\newcommand{\MNRAS}[3]{{ Mon. Not. Roy. Astron. Soc.} {\bf #1} (#2)  {#3}}
\newcommand{\PR}[3]{{ Phys. Rep. } {\bf #1} (#2)  {#3}}
\newcommand{\RMP}[3]{{ Rev. Mod. Phys. } {\bf #1} (#2)  {#3}}
\newcommand{\JPA}[3]{{ J. Phys. A: Math. Theor.} {\bf #1} (#2)  {#3}}
\newcommand{\ProgS}[3]{{ Prog. Theor. Phys. Supp.} {\bf #1} (#2)  {#3}}
\newcommand{\APJS}[3]{{ Astrophys. J. Supl.} {\bf #1} (#2)  {#3}}

\newcommand{\Prog}[3]{{ Prog. Theor. Phys.} {\bf #1}  (#2) {#3}}
\newcommand{\IJMPA}[3]{{ Int. J. of Mod. Phys. A} {\bf #1}  {(#2)} {#3}}
\newcommand{\IJMPD}[3]{{ Int. J. of Mod. Phys. D} {\bf #1}  {(#2)} {#3}}
\newcommand{\GRG}[3]{{ Gen. Rel. Grav.} {\bf #1}  {(#2)} {#3}}


\end{document}